\documentclass[a4paper,11pt,headings=big,DIV=12]{scrartcl}
\ifx\pdfoutput\undefined
\else
  \pdfoutput=1
\fi
\usepackage{amsmath,amssymb}
\usepackage{color,xcolor}
\definecolor{blue}{rgb}{0,0,0.5}
\usepackage[colorlinks,linkcolor=blue,urlcolor=blue,citecolor=blue]{hyperref}
\usepackage{graphicx}
\addtokomafont{disposition}{\rmfamily\boldmath}
\usepackage{cite}
\usepackage{xspace}
\usepackage{relsize}
\usepackage{bm}
\usepackage{multirow}
\usepackage{booktabs}
\usepackage[utf8]{inputenc}
\usepackage{arydshln}
\usepackage{csquotes}
\usepackage{bbold}

\definecolor{maroon}{cmyk}{0, 0.87, 0.68, 0.32}
\definecolor{halfgray}{gray}{0.55}
\definecolor{ipython_frame}{RGB}{207, 207, 207}
\definecolor{ipython_bg}{RGB}{255, 255, 255}
\definecolor{ipython_red}{RGB}{186, 33, 33}
\definecolor{ipython_green}{RGB}{0, 128, 0}
\definecolor{ipython_cyan}{RGB}{64, 128, 128}
\definecolor{ipython_purple}{RGB}{170, 34, 255}

\usepackage{listings}
\lstdefinelanguage{iPython}{
    morekeywords={access,and,break,class,continue,def,del,elif,else,except,exec,finally,for,from,global,if,import,in,is,lambda,not,or,pass,print,raise,return,try,while},%
    morekeywords=[2]{abs,all,any,basestring,bin,bool,bytearray,callable,chr,classmethod,cmp,compile,complex,delattr,dict,dir,divmod,enumerate,eval,execfile,file,filter,float,format,frozenset,getattr,globals,hasattr,hash,help,hex,id,input,int,isinstance,issubclass,iter,len,list,locals,long,map,max,memoryview,min,next,object,oct,open,ord,pow,property,range,raw_input,reduce,reload,repr,reversed,round,set,setattr,slice,sorted,staticmethod,str,sum,super,tuple,type,unichr,unicode,vars,xrange,zip,apply,buffer,coerce,intern},%
    sensitive=true,%
    morecomment=[l]\#,%
    morestring=[b]',%
    morestring=[b]",%
    morestring=[s]{'''}{'''},
    morestring=[s]{"""}{"""},
    morestring=[s]{r'}{'},
    morestring=[s]{r"}{"},%
    morestring=[s]{r'''}{'''},%
    morestring=[s]{r"""}{"""},%
    morestring=[s]{u'}{'},
    morestring=[s]{u"}{"},%
    morestring=[s]{u'''}{'''},%
    morestring=[s]{u"""}{"""},%
    literate=
    {á}{{\'a}}1 {é}{{\'e}}1 {í}{{\'i}}1 {ó}{{\'o}}1 {ú}{{\'u}}1
    {Á}{{\'A}}1 {É}{{\'E}}1 {Í}{{\'I}}1 {Ó}{{\'O}}1 {Ú}{{\'U}}1
    {à}{{\`a}}1 {è}{{\`e}}1 {ì}{{\`i}}1 {ò}{{\`o}}1 {ù}{{\`u}}1
    {À}{{\`A}}1 {È}{{\'E}}1 {Ì}{{\`I}}1 {Ò}{{\`O}}1 {Ù}{{\`U}}1
    {ä}{{\"a}}1 {ë}{{\"e}}1 {ï}{{\"i}}1 {ö}{{\"o}}1 {ü}{{\"u}}1
    {Ä}{{\"A}}1 {Ë}{{\"E}}1 {Ï}{{\"I}}1 {Ö}{{\"O}}1 {Ü}{{\"U}}1
    {â}{{\^a}}1 {ê}{{\^e}}1 {î}{{\^i}}1 {ô}{{\^o}}1 {û}{{\^u}}1
    {Â}{{\^A}}1 {Ê}{{\^E}}1 {Î}{{\^I}}1 {Ô}{{\^O}}1 {Û}{{\^U}}1
    {œ}{{\oe}}1 {Œ}{{\OE}}1 {æ}{{\ae}}1 {Æ}{{\AE}}1 {ß}{{\ss}}1
    {ç}{{\c c}}1 {Ç}{{\c C}}1 {ø}{{\o}}1 {å}{{\r a}}1 {Å}{{\r A}}1
    {€}{{\EUR}}1 {£}{{\pounds}}1
    {^}{{{\color{ipython_purple}\^{}}}}1
    {=}{{{\color{ipython_purple}=}}}1
    {+}{{{\color{ipython_purple}+}}}1
    {*}{{{\color{ipython_purple}$^\ast$}}}1
    {/}{{{\color{ipython_purple}/}}}1
    {+=}{{{+=}}}1
    {-=}{{{-=}}}1
    {*=}{{{$^\ast$=}}}1
    {/=}{{{/=}}}1,
    literate=
    *{-}{{{\color{ipython_purple}-}}}1
     {?}{{{\color{ipython_purple}?}}}1,
    identifierstyle=\color{black}\ttfamily,
    commentstyle=\color{ipython_cyan}\ttfamily,
    stringstyle=\color{ipython_red}\ttfamily,
    keepspaces=true,
    showspaces=false,
    showstringspaces=false,
    rulecolor=\color{ipython_frame},
    frame=single,
    frameround={t}{t}{t}{t},
    framexleftmargin=6mm,
    numbers=left,
    numberstyle=\tiny\color{halfgray},
    backgroundcolor=\color{ipython_bg},
    basicstyle=\footnotesize\ttfamily,
    keywordstyle=\color{ipython_green}\ttfamily,
    aboveskip=1.2em,
    belowskip=1.2em,
}

\newcommand{\wcxf}{\texttt{WCxf}}
\newcommand{\smelli}{\texttt{smelli}}
\newcommand{\wilson}{\texttt{wilson}}
\newcommand{\flavio}{\texttt{flavio}}
\newcommand{\ck}{\texttt{ClusterKinG}}

\newcommand{\SPheno}{\texttt{SPheno}\xspace}
\newcommand{\FlavorKit}{\texttt{FlavorKit}\xspace}
\newcommand{\FF}{\texttt{FormFlavor}\xspace}
\newcommand{\SMEFTsim}{\texttt{SMEFTsim}\xspace}

\newcommand{\SFR}{\texttt{SMEFT Feynman Rules}\xspace}

\newcommand{\pell}{\vec{p}_\ell^{\,\,*}}

\makeatletter
\def\adl@drawiv#1#2#3{%
        \hskip.5\tabcolsep
        \xleaders#3{#2.5\@tempdimb #1{1}#2.5\@tempdimb}%
                #2\z@ plus1fil minus1fil\relax
        \hskip.5\tabcolsep}
\newcommand{\cdashlinelr}[1]{%
  \noalign{\vskip\aboverulesep
           \global\let\@dashdrawstore\adl@draw
           \global\let\adl@draw\adl@drawiv}
  \cdashline{#1}
  \noalign{\global\let\adl@draw\@dashdrawstore
           \vskip\belowrulesep}}
\makeatother

\usepackage{rotating}

\allowdisplaybreaks

\DeclareOldFontCommand{\bf}{\normalfont\bfseries}{\mathbf}

\def \refeq#1{Eq.~(\ref{#1})}
\def \refsec#1{Sec.~\ref{#1}}

\def \reffig#1{figure~\ref{#1}}

\def \reftab#1{table~\ref{#1}}

\newcommand*{\clusterking}{\texttt{ClusterKinG}}
\usepackage{subcaption}
\begin{document}

\begin{flushleft}
\end{flushleft}

\vspace{-14mm}
\begin{flushright}
\end{flushright}

\vspace{0mm}

\begin{center}
{\LARGE\bfseries \boldmath
\vspace*{1.5cm}
Clustering of $\bar B\to D^{(*)}\tau^-\bar\nu_\tau$ kinematic distributions with ClusterKinG
}\\[0.8 cm]
{\textsc{
Jason Aebischer$^a$,
Thomas Kuhr$^{b,c}$,
Kilian Lieret$^{b,c}$
}\\[0.5 cm]
\small
$^a$ Excellence Cluster Universe, Garching, Germany\\
$^b$ Excellence Cluster Origins, Garching, Germany\\
$^c$ Ludwig Maximilian University, Munich, Germany
}
\\[0.5 cm]
\footnotesize
E-Mail:
\texttt{jason.aebischer@tum.de},
\texttt{thomas.kuhr@lmu.de},
\texttt{kilian.lieret@lmu.de}
\\[0.2 cm]
\end{center}

\bigskip

\begin{abstract}\noindent
New Physics can manifest itself in kinematic distributions of particle decays.
The parameter space defining the shape of such distributions can be large which is challenging for both theoretical and experimental studies.
Using clustering algorithms, the parameter space can however be dissected into subsets (clusters) which correspond to similar kinematic distributions.
Clusters can then be represented by benchmark points, which allow for less involved studies and a concise presentation of the results.
We demonstrate this concept using the Python package \ck{}, an easy to use framework for the clustering of distributions that particularly aims to make these techniques more accessible in a High Energy Physics context.
As an example we consider $\bar B\to D^{(*)}\tau^-\bar\nu_\tau$ distributions and discuss various clustering methods and possible implications for future experimental analyses.
\end{abstract}
\newpage

\setcounter{tocdepth}{2}
\tableofcontents

\newpage

%
%
%
%
%
\section{Introduction}
New Physics (NP) contributions can influence the kinematic distributions of particle decays.
While this opens up possibilities to find and determine the nature of NP, it can also be a nuisance for experimental studies, because most measurements require assumptions on certain kinematic distributions, e.g.\ to distinguish signal from background or to determine efficiencies.

For example, as was shown in \cite{Lees:2012xj}, assuming a two-Higgs-doublet model of type II changes the experimental measurement of $R(D^{(*)})$, because discriminating between signal and background requires assumptions on the kinematic shapes of the signal, background and normalisation modes. Such kinematic shapes are in general determined from Monte Carlo simulations.

Thus, many experimental measurements are model-dependent and are often only conducted under the assumption of the Standard Model (SM).
Discrepancies between the SM prediction and the measured values are a good indication for NP. However comparing different NP models based on their predicted results has to be taken with a grain of salt, because the measurements themselves are model-dependent.

A further complication for both theoretical and experimental studies is the high dimensionality of the parameter space of typical NP models.
If experimentalists wish to publish model-independent results, the studies have to be repeated for a large sample of parameter points.
This can be computationally very expensive.
Furthermore numerical results and many visualisations can only be shown for specific (often arbitrary) parameter points, leaving their representative character unknown.

A possibility to reduce the complexity of this problem is to identify regions in the parameter space which lead to similar kinematic distributions.
These regions can be found using clustering algorithms.
From each cluster, a most representative point (\emph{benchmark point}, BP) can then be chosen.

Experimental studies can focus on these BPs, thereby reducing the multi-dimensional problem to a small number of BPs to be considered.
The results can be presented for each BP, allowing for a clear-cut numerical result and simpler visualisations.

Such a strategy has been employed for the first time in the context of Higgs boson pair production in \cite{Carvalho:2015ttv,Carvalho:2016rys,Carvalho:2017vnu}. An effective field theory (EFT) approach has been adopted to parametrise the five-dimensional parameter space of anomalous Higgs couplings.
It has been shown in \cite{Carvalho:2015ttv}, that for current and future LHC searches a total of 12 clusters gives a reasonable approximation of the considered parameter space.
In \cite{Carvalho:2016rys} several clusters were subjected to experimental limits from the CMS collaboration \cite{Khachatryan:2016sey}. Finally, in \cite{Carvalho:2017vnu} a method to extend the experimental sensitivity from the BPs to the other cluster members is discussed.

In recent years substantial progress has been made in the EFT description of the SM in form of the SM Effective Theory (SMEFT) \cite{Grzadkowski:2010es} and the Weak Effective Theory (WET).
The calculation of the complete one-loop SMEFT renormalization group equations (RGEs) \cite{Jenkins:2013zja, Jenkins:2013wua, Alonso:2013hga, Celis:2017hod},
the complete tree-level and one-loop matching from SMEFT onto WET \cite{Aebischer:2015fzz, Jenkins:2017jig,Dekens:2019ept} and the complete one-loop QCD and QED RGEs within WET \cite{Aebischer:2017gaw, Jenkins:2017dyc} allow for a general NP analysis of low-energy observables.

Various tools are dedicated to the study of Wilson coefficients above and below the EW scale.
They include the Match(runn)ers \texttt{DsixTools} \cite{Celis:2017hod}, \texttt{MatchingTools} \cite{Criado:2017khh} and \wilson \cite{Aebischer:2018bkb}, the Wilson coefficient exchange format \wcxf\cite{Aebischer:2017ugx}, the fitting tool \smelli\cite{Aebischer:2018iyb}, basis codes like \texttt{BasisGen} \cite{Criado:2019ugp}, \texttt{DEFT} \cite{Gripaios:2018zrz} and \texttt{abc-eft} \cite{Brivio:2019irc}, the observable calculator \flavio \cite{Straub:2018kue}, \FlavorKit~\cite{Porod:2014xia}, \texttt{SuperIso} \cite{Mahmoudi:2007vz}, \SPheno~\cite{Porod:2003um,Porod:2011nf}, \FF \cite{Evans:2016lzo}, as well as packages related to SMEFT, such as \SFR~\cite{Dedes:2019uzs} and \SMEFTsim~\cite{Brivio:2017btx}.

However public tools to cluster the phenomenology of NP operators in a systematic way are still missing so far.

To fill this gap we have written the Python package \ck{} (\underline{Cluster}ing of \underline{Kin}ematic \underline Graphs), which aims to make clustering techniques more easily accessible in the context of EFTs and High Energy Physics in general.
Despite this motivation, \ck{} is a general tool that can be applied to a diverse set of problems, even outside of physics.

In this article, we use this package to cluster kinematic distributions of $\bar B\to D^{(*)}\tau^-\bar\nu_\tau$ decays.
These decays are of particular interest in view of the current $B$ anomalies \cite{Altmannshofer:2017poe,Murgui:2019czp,Shi:2019gxi,Becirevic:2019tpx,Gomez:2019xfw,Alguero:2019ptt,Ciuchini:2019usw,Datta:2019zca,Aebischer:2019mlg}.

The rest of the paper is organized as follows: In \refsec{sec:cluster} we discuss the clustering method in general terms. In \refsec{sec:CK} we present for the first time the \ck{} package and describe its features. In \refsec{sec:numerics} we apply the clustering method to kinematic distributions of the decays $\bar B\to D^{(*)}\,\tau^-(\to \ell\, \bar\nu_\ell\, \nu_\tau)\bar\nu_\tau$ and perform various consistency tests.
Finally we conclude in \refsec{sec:conclusions}.

\section{Clustering}\label{sec:cluster}
In this section we discuss the different steps involved in our clustering approach.
As a first step a suitable observable and the corresponding model of interest have to be chosen.
After establishing the underlying parameter space the following steps will be performed:
\begin{itemize}
\item Sampling the parameter space of the process.
\item Computing the kinematic distributions in question.
\item Choosing a metric to measure differences between the kinematic distributions.
\item Applying a suited clustering algorithm on the parameter space.
\item Computing the BPs representing each of the clusters.
\end{itemize}
The above steps are explained in the following subsections.

\subsection{Sampling of the parameter space}
As discussed in the introduction, typical NP models depend on several parameters.
Theoretical considerations (such as symmetry arguments) can often be used to limit the study to a subset of these parameters, thereby reducing the dimensionality of the problem.
The considered range of these parameters can be motivated by existing exclusion limits or from theory.

From this subset of the original parameter space, a set of points, \emph{sample points}, are taken and used for the rest of the analysis.
While large numbers of sample points will make the following steps more computationally expensive, it is important that the sampling is fine enough to accurately represent the whole parameter space.

\ck{} allows for an arbitrary selection of sampling points. The examples presented in this paper use a uniform grid in the parameter space for simplicity. In order to limit the number of required points (and thereby computing time), it is also planned to implement adaptive sampling techniques in the future: After an initial run with a coarse grid, regions with large variations in the kinematic distributions are identified and sampled in a finer grid. If needed, this procedure can then be applied several times.

\subsection{Kinematic distributions}\label{subsec:kind}
For every sample point, the corresponding kinematic distribution needs to be computed.
If analytic expressions of the observable in terms of the parameters are available, this task can be achieved by explicit evaluation of the formulae.
Otherwise Monte Carlo (MC) simulations have to be used to generate the distributions.
Since the generation of MC samples is in general resource-intensive, reweighting techniques can be used to generate samples corresponding to different parameter points from existing samples. For semileptonic $B$ decays such methods are already implemented in the \texttt{HAMMER} tool \cite{Bernlochner:2020tfi,Duell:2016maj}.

In this article we only consider \emph{binned} observables and our kinematic distributions are thus histograms.

\subsection{Metric}

The objective of the clustering procedure is to partition the parameter space in such a way that parameter points generating similar kinematic distributions are grouped into the same cluster.
For this, the \enquote{similarity} between kinematic distributions has to be quantified by a metric\footnote{Now and in the following, the term \enquote{metric} is used in a rather loose way, emphasizing the intuition of a distance measure while not necessarily fulfilling all requirements to be a metric in the mathematical sense.} in the space of kinematic distributions.

The choice of this metric follows from the interpretation of the distributions as potential measurements.
As such, the metric of choice should give more weight to differences in bins that are associated with low expected experimental uncertainties, while being less sensitive to differences in bins of less experimental significance.
Having an estimate of experimental uncertainties is also useful when deciding the number of clusters and benchmark points to present: Sufficiently many to cover the whole variety of distributions that lead to distinguishable experimental results, but not arbitrarily many.
In this way, the number of clusters then also serves as an estimate for the sensitivity of an distribution to NP parameters.

A common choice for a metric measuring the similarity between binned distributions is a two-sample test statistic such as the Kolmogorov-Smirnov test \cite{kolmogorov_1951}, the Anderson-Darling test \cite{10.1093/biomet/63.1.161} or the $\chi^2$ test.

In \cite{Carvalho:2015ttv} a binned log likelihood test statistic is used to distinguish between two distributions.
This likelihood ratio is obtained by taking the ratio of the binomial distribution of the two individual samples and the binomial distribution where both samples are assumed to be equal.
The log of this ratio can be shown to be $\chi^2$-distributed up to a minus sign \cite{wilks1938}.
By basing the test statistic on binomial distributions, the metric incorporates the statistical significance of the different bins.

In this paper, we use a $\chi^2$ test operating on normalised distributions with uncertainties.
As the distributions are not measured but generated as described in \refsec{subsec:kind}, an uncertainty estimate is applied to them, consisting of configurable statistical uncertainties and relative and absolute systematic uncertainties that can be correlated between bins.

Let $n_{ki}$ be the bin contents of two histograms $H_k$ ($k=1,2$).
Our null hypothesis is that the bin contents of the histograms are drawn from two distributions with identical means.
We assume that the $n_{ki}$ are distributed according to a multivariate normal distribution with covariance matrices $\Sigma_k=\bigl(\operatorname {Cov}(n_{ki}, n_{kj})\bigr)_{ij}$.
We denote the corresponding normalisations as $N_k = \sum_{i=1}^N n_{ki}$ and define $\Delta_i = \frac{n_{1i}}{N_1} - \frac{n_{2i}}{N_2}$ and  $\Sigma=\frac{\Sigma_1}{N_1^2} + \frac{\Sigma_2}{N_2^2}$.
Our $\chi^2$ measure is then given by
\begin{equation}
\chi^2(H_1, H_2) = \sum_{i,j=1}^N \Delta_i (\Sigma^{-1})_{ij} \Delta_j.
\label{chi2_new}
\end{equation}
Under the null hypothesis $\chi^2(H_1,H_2)$ approximates a $\chi^2$ distribution with $N-1$ degrees of freedom, henceforth denoted as $\chi^2_{N-1}$.

It should be highlighted that this approximation can break down if the uncertainties $\sigma_{ki}$ are very imbalanced, though this does not usually happen if the uncertainties are dominated by Poisson uncertainties.
App.~\ref{appendix:validation} describes toy studies that were used to validate the statistical treatment for all results shown in  \refsec{sec:numerics}.

\begin{figure}
	\centering
	\includegraphics[scale=0.65]{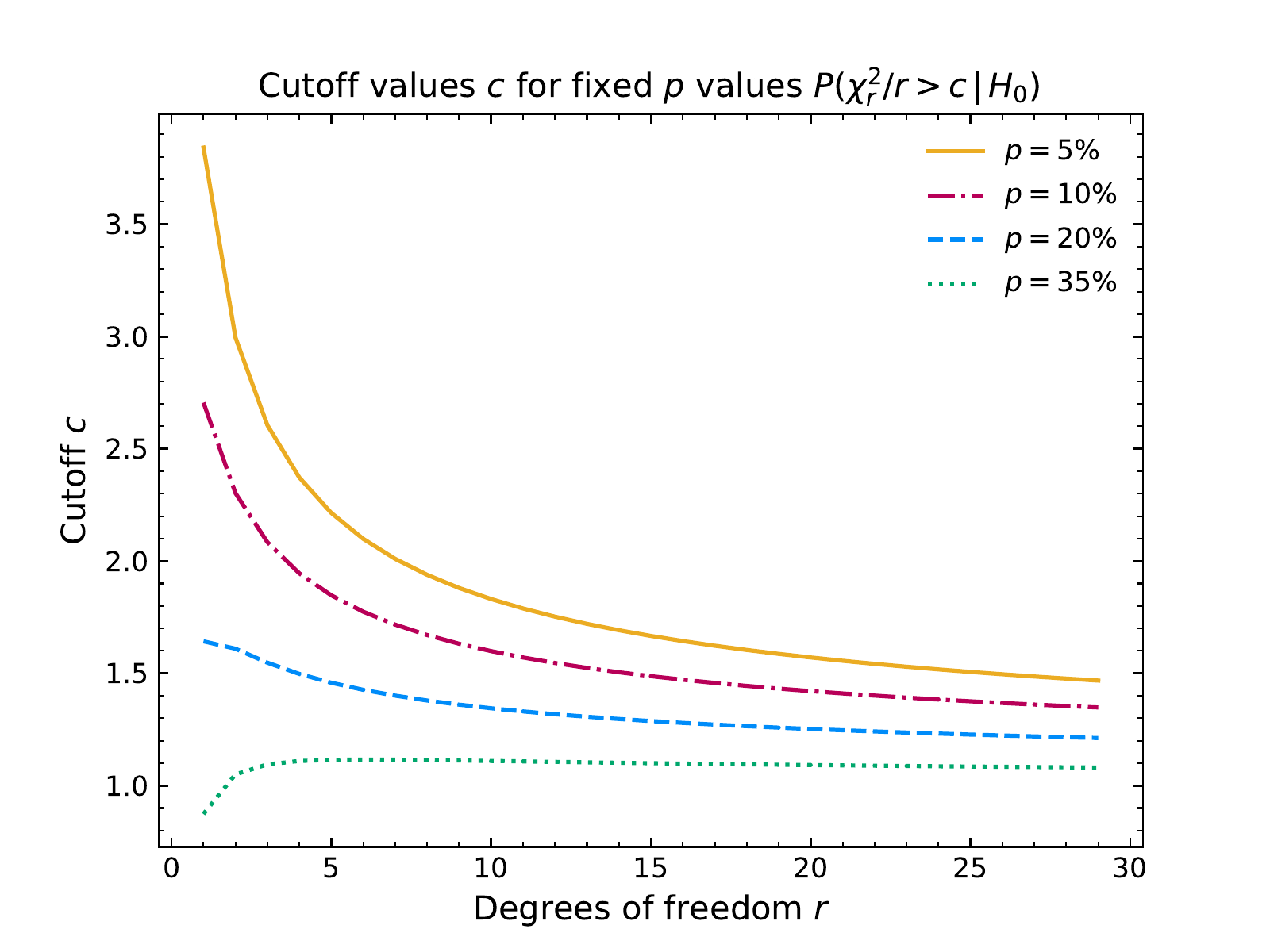}
	\caption{$p$ values and corresponding cut-off values.}
	\label{pvalue_cutoff}
\end{figure}

In the following we call two distributions distinguishable if their $\chi^2$-distance $\chi^2(H_1, H_2)/(N-1)$ is larger than $1.125$, corresponding to a $p$-value of $34\%$ for $N=9$.\footnote{the unround number of $1.125$ was chosen for consistency with a previous version of this paper which incorrectly assumed $N$ degrees of freedom.}
Figure \ref{pvalue_cutoff} shows the relationship between this \emph{cutoff value}, the $p$-value and the number of bins.
This loose definition of distinguishability is conservative in the sense that it will lead to more clusters than a stricter criterion.
The metric between the kinematic distributions gives rise to a metric $d$ acting directly on the parameter space. In our case:
\begin{equation}\label{eq:metric}
    d(c_1, c_2)\equiv  \chi^2(H_1, H_2)/(N-1)\,,
\end{equation}
for two sample points $c_{1,2}$ and their respective histograms $H_{1,2}$.

\subsection{Clustering algorithm}

In general a data set can be either clustered \emph{hierarchically} or \emph{partitionally}.
Partitional clustering methods \cite{Hartigan:105051} such as for example K-means algorithms \cite{macqueen1967} only perform one single partition of the data.
Furthermore the number of resulting clusters has to be chosen beforehand as an input parameter.
In the following we focus on hierarchical clustering methods \cite{Kaufman1990FindingGI}.

Hierarchical clustering algorithms group a given set of data in a nested sequence of clusters.
Two approaches are common:
Bottom-up (or agglomerative) algorithms successively merge clusters to form larger clusters, whereas top-down algorithms successively split clusters into smaller ones.
In both cases, a stopping criterion is needed to avoid a trivial result.
In our analysis we will employ the agglomerative method with the following steps:

\begin{enumerate}
\item Associate each sample point to one cluster containing only this element.
\item Merge the nearest two clusters according to a metric $D$.
\item Repeat step 2 until the stopping criterion is fulfilled.
\end{enumerate}

Note that the metric $D$ in step 2 is not between points in the parameter space, but between subsets (existing clusters) of this space.
It makes sense to base $D$ on the metric $d$ introduced in \refeq{eq:metric}.
Two canonical choices for the inter-cluster metric $D$ are
\begin{align}
    D_\infty(C_1, C_2)\equiv  \max_{c_1\in C_1,\, c_2\in C_2} d(c_1, c_2)\,,
    \quad
    D_1(C_1, C_2)\equiv  \frac 1{|C_1||C_2|}\sum_{c_1\in C_1,\, c_2\in C_2} d(c_1, c_2)\,,
    \label{eq:metric_D}
\end{align}
where $C_1$ and $C_2$ are clusters and $|C_{1,2}|$ denote their number of elements respectively.

While \cite{Carvalho:2015ttv} uses the \enquote*{average} metric $D_1$, we will employ the $D_\infty$ metric in our analysis.
This choice is more conservative in the sense that it usually leads to a larger number of clusters than in the case of $D_1$, simply because $D_\infty\geq D_1$.

The stopping criterion is chosen to be $D_\infty(C_1, C_2) > 1.125$ for all pairs of clusters.
This means that if we merge two clusters the resulting larger cluster does not contain any two distinguishable points.
As a consequence all clusters of our final result contain only indistinguishable sample points.
This is not the case for the \enquote*{average} metric in general.

\subsection{Benchmark points}
After the application of the clustering algorithm BPs have to be determined for all of the resulting clusters.
A BP is a cluster element $c$ of a cluster $C$, which is chosen as a representative of that particular cluster.
Usually it is taken to be the parameter point that minimizes a certain figure of merit, which is commonly based on the metric $d$ of \refeq{eq:metric}. Examples are:
\begin{align}\label{eq:FOM}
    f_1(c, C) \equiv \frac 1{|C|}\sum_{c_i\in C} d(c, c_i),
    \quad f_2(c, C) \equiv \sqrt{\sum_{c_i\in C} d(c, c_i)^2},
    \quad f_\infty(c, C) &\equiv \max_{c_i\in C} d(c, c_i),
\end{align}
which differ in their responsiveness to outliers in the data.
The BPs are the key elements of the cluster analysis.
They are determined to simplify experimental analyses, which can then lay their focus only on a finite set of BPs instead of the entire parameter space.

\section{The ClusterKinG package}\label{sec:CK}
\ck{} is publicly developed as open source under MIT license on \href{https://github.com/clusterking}{\texttt{github.com/clusterking}}. The package aims for ease of use, while making it easy for users to manipulate and extend functionality.

The basic building blocks of \ck{} are worker classes: After initialisation, a set of methods can be called for further configuration, before calling a \texttt{run} method, that performs the desired action.
By subclassing these worker classes, the functionality can be extended.
Furthermore, these worker classes can be passed on to other worker classes to perform e.g.
stability checks that require repetitive calling with slight parameter variations

To demonstrate the ease of use, a fully working example of code to generate clusters, benchmark points and plots similar to the ones presented in \refsec{sec:numerics} is shown in App.~\ref{app:ck}.

\medskip
A typical \clusterking{} workflow consists of the following steps:
\begin{enumerate}
   \item
   Initialise a \texttt{Data} object: this is what the following worker classes read from and write to.

   Internally this class holds a \texttt{pandas} \texttt{DataFrame} \cite{mckinney-proc-scipy-2010} that contains the bin contents of the kinematic distributions for each sample point.
   Additional information (\emph{metadata}) is saved as a nested dictionary.
   Each of the following steps will add information to the metadata, such that any output file contains a description of how it was generated.

   Data objects can be saved to and loaded from a single output file in SQL format, allowing to save all data in a single output file of comparably small size.
   Exports of the data to other formats (CSV, XLS, JSON, ...) are easily available.
   \item Adding uncertainties (optional).
   After the distributions have been generated, an experimental uncertainty estimate can be added to them.
   Typically this consists of
   \begin{itemize}
      \item Statistical uncertainties, modeled as Poisson errors (and thereby dependent on the actual bin content for the distribution for each sample point).
      \item Systematic uncertainties which can be given in absolute form or relative to the bin content for all sample points at once.
   \end{itemize}
   To save memory space and improve performance, the uncertainties on the distributions are only calculated when needed and are not saved to the Data object, but only the procedure of \emph{how} they are calculated.
   This also means that it is very easy and fast to check the dependency of the clustering results on the uncertainties.
   \item
   Scanning: (Kinematic) distributions are generated for the chosen sample points.
   This is done by a \texttt{Scanner} object or for convenient clustering in the space of Wilson coefficients its \texttt{WilsonScanner} subclass.

   Generally this requires three steps:
   \begin{enumerate}
      \item Providing a function to generate kinematic distributions: Any Python function can be used, in particular any observable from the \texttt{flavio} \cite{Straub:2018kue} package.
      \item Defining the sample points, usually in the form of an equidistant grid in the parameter dimensions (though more fine-tunable methods are available).
      The input of \texttt{WilsonScanner} is given in the common \texttt{WCxf} format defined in \cite{Aebischer:2017ugx}.
      The Wilson coefficients are specified at a certain scale and in a particular EFT and basis.
      \item Running the scanning process. \clusterking{} supports multiprocessing to speed up calculations.
    \end{enumerate}
    The resulting distributions and the spread of the bin contents among the sample points are visualised by calling some of the plot methods of the \texttt{Data} object.
    \item Clustering: Different clustering algorithms correspond to different subclasses of the \texttt{Cluster} class.
    Hierarchical clustering is implemented in the \texttt{HierarchyCluster} worker which internally uses algorithms of the \texttt{scipy} library \cite{scipy} and can be used with a range of different metrics ($p$-metrics, $\chi^2$ metric, user defined functions).
    \item Selection of benchmark points is done by the \texttt{Benchmark} class, which can be configured (and subclassed) to allow for different strategies.
\end{enumerate}
Besides visualisation, the data class provides simple methods to find the closest benchmark or sampling points given a point in parameter space, among others.

The technical documentation of this package is available at \href{https://clusterking.readthedocs.io}{\texttt{clusterking.readthedocs.io}} and multiple usage examples are provided in the form of Jupyter notebooks in the main repository.

\section{Clustering of \texorpdfstring{$B\to D^{(*)}\tau\bar\nu$}{B -> D(*) tau nu} distributions}\label{sec:numerics}
\subsection{Setup}
In this subsection we describe the setup used for our numerical analysis. Motivated by current $B$ anomalies, we perform a clustering analysis on various $\bar B\to D^{(*)}\,\tau^-\bar\nu_\tau$ kinematic distributions.
Such $b\to c$ transitions are described by the following effective Lagrangian:
\begin{equation}\label{eq:Lbc}
  \mathcal{L}_\text{eff} = -\frac{4G_F}{\sqrt{2}}V_{cb}\left[C_{VL}O_{VL}+C_{VR}O_{VR}+C_{SL}O_{SL}+C_{SR}O_{SR}+C_{T}O_{T}\right]+\text{h.c.}\,,
\end{equation}
with the CKM element $V_{cb}$, the Fermi coupling constant $G_F$ and the effective operators
\begin{align}
  O_{VL}&= (\bar c \gamma_\mu P_L b)(\bar \tau \gamma^\mu P_L \nu_\tau)\,, &  O_{VR}&= (\bar c \gamma_\mu P_R \,b)(\bar \tau \gamma^\mu P_L \nu_\tau)\,,\notag\\
  O_{SL}&= (\bar c P_L b)(\bar \tau  P_L \nu_\tau)\,,& O_{SR}&= (\bar c  P_R b)(\bar \tau  P_L \nu_\tau)\,, \\
  O_{T}&= (\bar c \sigma_{\mu\nu} P_L b)(\bar \tau \sigma^{\mu\nu} P_L \nu_\tau)\,. &\notag
\end{align}
We use the notation $P_{R,L}=\frac{1}{2}(\mathbb{1}\pm\gamma_5)$, $\sigma^{\mu\nu}=\frac{\mathrm i}{2}[\gamma^\mu,\gamma^\nu]$ and $b,c,\tau,\nu_\tau$ for the quark and lepton fields.
The Wilson coefficients in \refeq{eq:Lbc} are in general complex quantities.
For our analysis we will assume the CP conserving limit, i.e.\ real Wilson coefficients.
This is a common assumption\footnote{Studies with $CP$-violating contributions are analysed in \cite{Celis:2016azn
,Blanke:2018yud,Bhattacharya:2018kig,Becirevic:2018afm,Bhattacharya:2019olg}.} (see f.e. \cite{Murgui:2019czp,Shi:2019gxi}) and is mainly chosen for simplicity.
Furthermore, for presentational reasons we study three out of the five Wilson coefficients and choose one for each Dirac structure, namely:
\begin{equation}\label{eq:WCs}
  C_{VL},\, C_{SL},\, C_{T}\,.
\end{equation}
We will assume values for the first two Wilson coefficients in the interval $[-0.5, 0.5]$.
The tensor operator is constrained \cite{Aebischer:2018iyb} from the longitudinal polarization fraction $F_L$ in $B\to D^{0*}\tau\nu$ \cite{Adamczyk:2019wyt} and we choose its Wilson coefficient to be in the interval $[-0.1, 0.1]$.
For our analysis we have chosen an equidistant grid of 1000 sample points in the three-dimensional parameter space, where each of the Wilson coefficients lies within the aforementioned intervals.

The clustering is performed using the \ck{} package. As mentioned in \refsec{sec:cluster} we use the hierarchical clustering algorithm together with the $\chi^2$-metric defined in \refsec{sec:cluster}.
The stopping criterion is chosen such that the $\chi^2$ distance between all distributions within the same cluster is $\leq 1.125$, meaning that they are indistinguishable experimentally.
Finally, the BPs are obtained by adopting the figure of merit $f_1$ from \refeq{eq:FOM}.

The complete code that has been used for the generation of the results and plots below is provided in the \texttt{example} directory of the \texttt{ClusterKinG} repository together with usage instructions.
\subsection{Results}
\subsubsection{$\cos(\theta_\tau)$-distribution of $\mathrm{BR}(B\to D^{0*} \tau\nu)$}
As a first example we consider nine bins of the $\cos(\theta_\tau)$-distribution of the branching ratio $\text{BR}(B\to D^{0*} \tau\nu)$, where $\theta_\tau$ denotes the angle between the $\tau$ and the $B$ meson in the dilepton mass frame.
The kinematic distributions are generated using the \texttt{flavio} \cite{Straub:2018kue} package.
With an assumed systematic uncertainty of 1\% and statistical uncertainties corresponding to a yield of $700$ events, our clustering procedure leads to a total of six clusters and their corresponding BPs.

\begin{figure}[tbp]
\centering
\includegraphics[trim=0 2.1cm 0 0, width=1.1\textwidth]{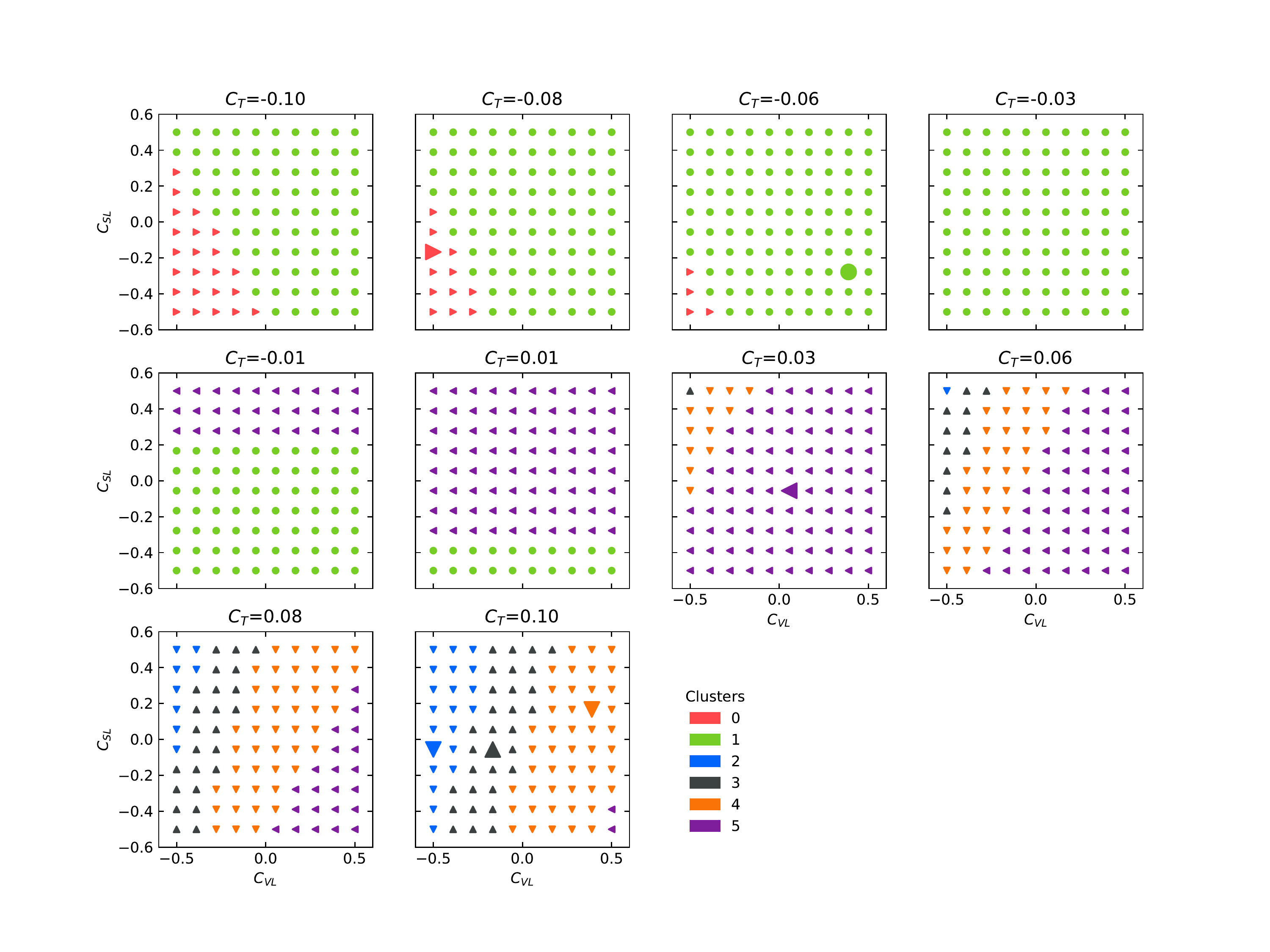}
\caption{Two-dimensional cuts of the clustered parameter space resulting from $\mathrm d\text{BR}(B\to D^{0*}\tau \nu)/\mathrm d(\cos(\theta_\tau))$ distributions is shown. The parameter space is spanned by the three Wilson coefficients $C_{VL},\,C_{SL}$ and $C_T$, varied with their respective ranges. Six different clusters are found in our approach, which are indicated with different markers and colours. BPs are given in boldface.}
\label{fig:cosl_clust2D}
\end{figure}

The clustered parameter space is shown as two dimensional cuts in the $C_T$-direction in \reffig{fig:cosl_clust2D} and numeric values for the BPs are reported in \reftab{tab:boxplot_BPs}.
As can be seen in \reffig{fig:cosl_clust2D}, the parameter space exhibits a strong cluster variation in the direction of $C_T$.
This fact is not surprising considering the explicit dependence of the kinematic distribution and agrees with the findings in \cite{Jung:2018lfu}, where this \enquote{flat term} observable has been proposed in the context of charged $b\to c$ transitions involving light leptons.

\begin{figure}[tbp]
\centering
\includegraphics[width=0.8\textwidth]{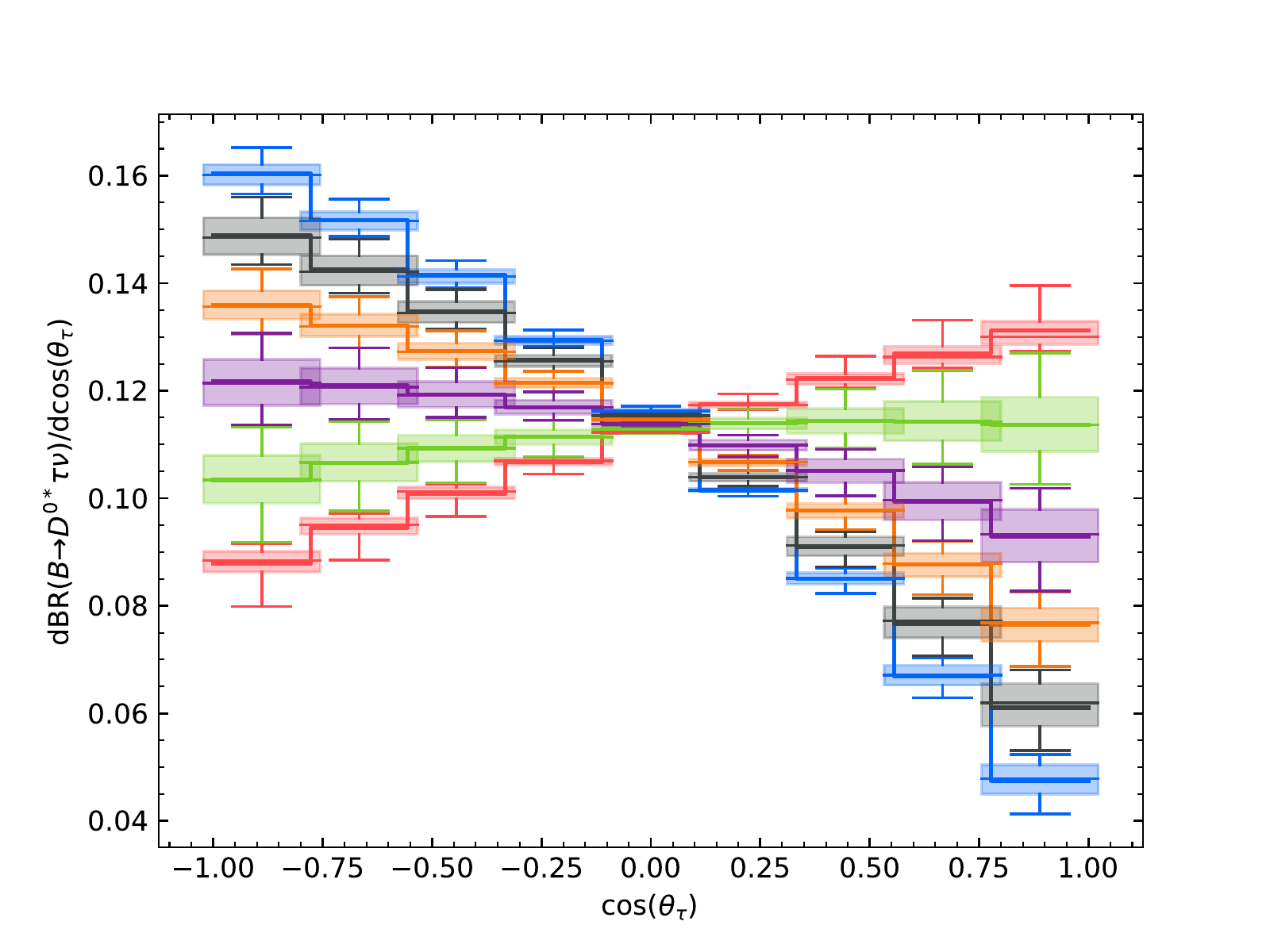}
\caption{The distributions of the observable $\mathrm d\text{BR}(B\to D^{0*}\tau \nu)/\mathrm d(\cos(\theta_\tau))$ for the six different clusters (with colours matching these of \reffig{fig:cosl_clust2D}).
The histograms corresponding to the the BPs are shown as solid lines. The boxes extend from the upper to the lower quartile of the distribution of the bin contents within a cluster and a horizontal line indicates the median.
Whiskers are used to further indicate the span of the data, covering six times the interquartile range.
Points beyond this range are plotted as individual points (outliers). }
\label{fig:cosl_box}
\end{figure}

The distributions corresponding to the sample points are visualised as a box plot in \reffig{fig:cosl_box}.
As expected, different clusters correspond to significantly different distributions. Furthermore, the distributions of the benchmark points are similar to the distributions given by the means of the bin contents of all distributions of the corresponding cluster.
\begin{table}[tbp]
    \centering
    \begin{tabular}{crrr}
    \toprule
     BP & $C_{VL}$ &  $C_{SL}$ &  $C_T$ \\
    \midrule
0 &         $-0.50$ &         $-0.17$ &        $-0.08$ \\
1 &          $0.39$ &         $-0.28$ &        $-0.06$ \\
2 &         $-0.50$ &         $-0.06$ &         $0.10$ \\
3 &         $-0.17$ &         $-0.06$ &         $0.10$ \\
4 &          $0.39$ &          $0.17$ &         $0.10$ \\
5 &          $0.06$ &         $-0.06$ &         $0.03$ \\
    \bottomrule
    \end{tabular}
    \caption{List of benchmark points for the distribution $\mathrm d\text{BR}(B\to D^{0*} \tau\nu)/\mathrm d (\cos(\theta_\tau))$ obtained from the clustering procedure given in terms of the left-handed vector, scalar and tensor Wilson coefficients $C_{VL}, C_{SL}$ and $C_T$.}
    \label{tab:boxplot_BPs}
\end{table}
\subsubsection{$\cos(\theta_V)$-distribution of $\mathrm{BR}(B\to D^{0*} \tau\nu)$}
\begin{figure}[tbp]
\centering
\includegraphics[trim=0 0cm 5cm 1.9cm, width=0.9\textwidth]{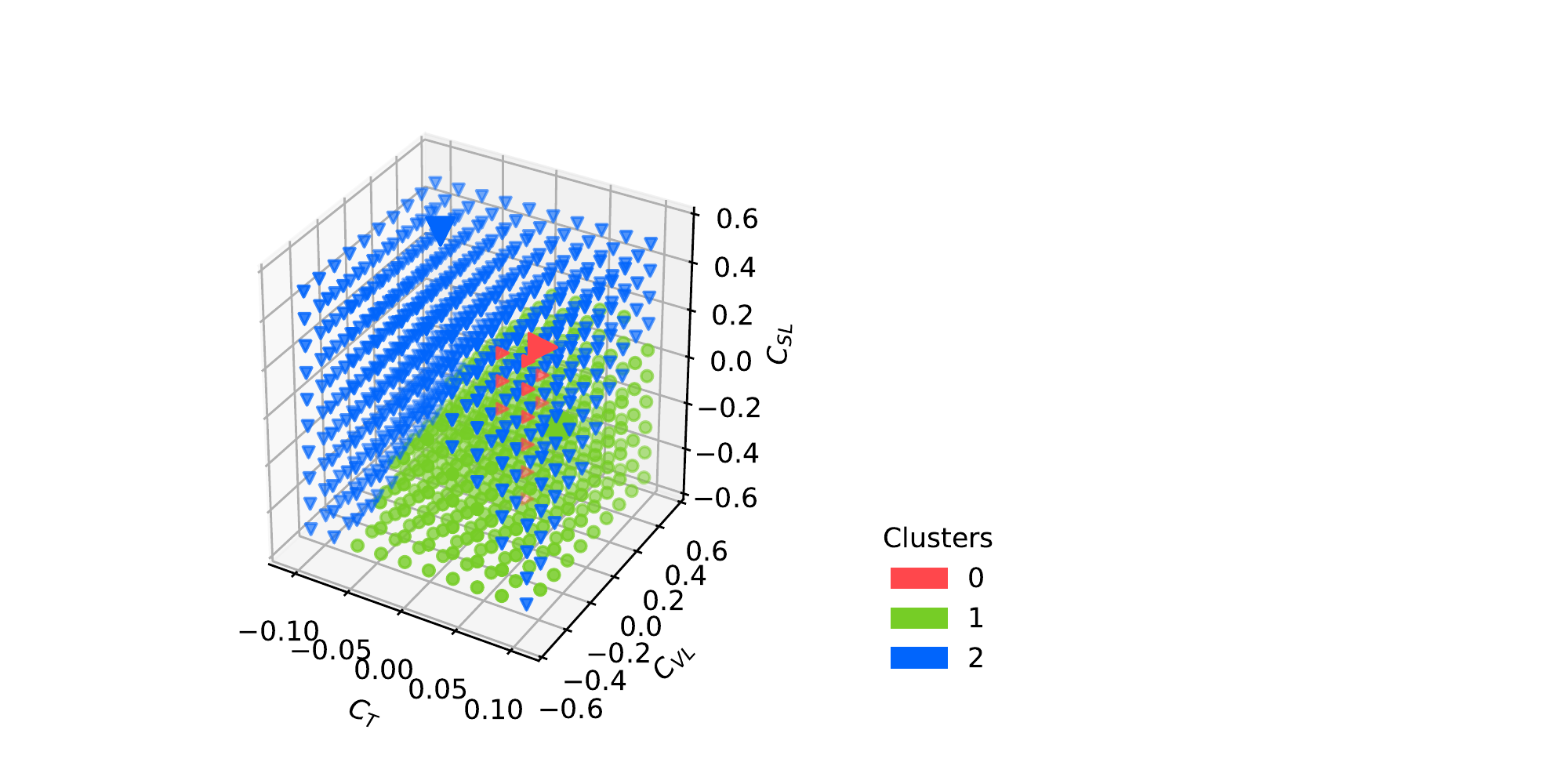}
\caption{The clustered three-dimensional parameter space resulting from $\mathrm d\text{BR}(B\to D^{0*}\tau \nu)/\mathrm d(\cos(\theta_V))$ distributions is shown. The parameter space is spanned by the three Wilson coefficients $C_{VL},\,C_{SL}$ and $C_T$, varied with their respective ranges. Three different clusters are found in our approach, which are indicated with different markers and colours. BPs are given in boldface.}
\label{fig:cosV_3D}
\end{figure}

As a second example we consider the $\cos(\theta_V)$-distribution of the process $B\to D^{0*} \tau\nu$.
Here $\theta_V$ denotes the angle between $D^{0*}$ and the $B$ meson.
The kinematic distributions for this process are again generated using \texttt{flavio} \cite{Straub:2018kue}.
Assuming a signal yield of $700$ events and a relative systematic uncertainty of 1\%, the clustering procedure leads to a total of three clusters.
The clustered three-dimensional sample space is shown in \reffig{fig:cosV_3D} and $C_T$ can again be identified to be the most influential Wilson coefficient.
A large subset of the parameter space belongs to cluster 2 (blue), whereas only a few sample points are contained in the first cluster (red), which is found at the edges of the sample space.
The three BPs are reported in \reftab{tab:lineplot_BPs}.

\begin{figure}[tbp]
\centering
\includegraphics[width=0.8\textwidth]{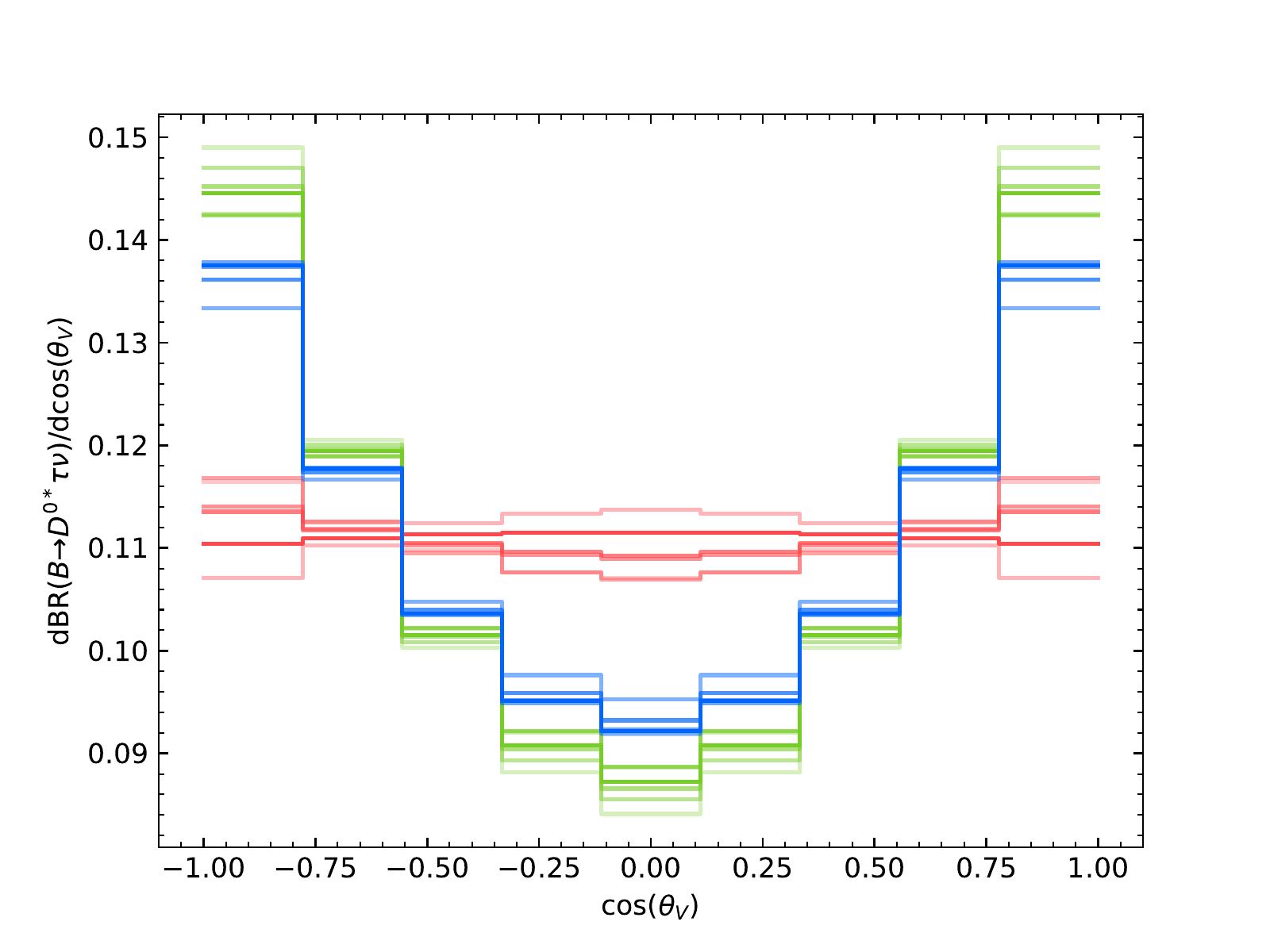}
\caption{$\cos(\theta_V)$-distributions of the decay $\bar B\to D^{0*}\tau^-\bar\nu_\tau$. corresponding to three sample points and the BPs.
Distributions corresponding to the same cluster are shown in the same colour (matching the colour scheme of \reffig{fig:cosV_3D}), with the sample distributions faded out slightly.}
\label{fig:cosV_dist}
\end{figure}

Finally we show several example distributions for each cluster together with the BP distributions in \reffig{fig:cosV_dist}.
While the distinction between the red cluster and the other two clusters is very clear, the kinematic distributions of the blue and green clusters are more similar.

Compared to $\theta_\tau$, less clusters are found, but the shapes of the clusters are different from the previous ones.
The two observables can therefore be considered complementary in their respective sensitivity to NP models.
\begin{table}[tbp]
  \centering
\begin{tabular}{crrr}
\toprule
 BP & $C_{VL}$ &  $C_{SL}$ &  $C_T$ \\
\midrule
0 &         $-0.39$ &          $0.50$ &         $0.10$ \\
1 &          $0.17$ &         $-0.17$ &         $0.06$ \\
2 &          $0.17$ &          $0.50$ &        $-0.06$ \\
\bottomrule
\end{tabular}
\caption{List of benchmark points for the distribution $\mathrm d\text{BR}(B\to D^{0*} \tau\nu)/\mathrm d (\cos(\theta_V))$ obtained from the clustering procedure given in terms of the left-handed vector, scalar and tensor Wilson coefficients $C_{VL}, C_{SL}$ and $C_T$.}
\label{tab:lineplot_BPs}
\end{table}
\subsubsection{$q^2$-distribution of $\text{BR}(B\to D^0 \tau\nu)$}
\begin{figure}[tbp]
\centering
\includegraphics[width=0.8\textwidth]{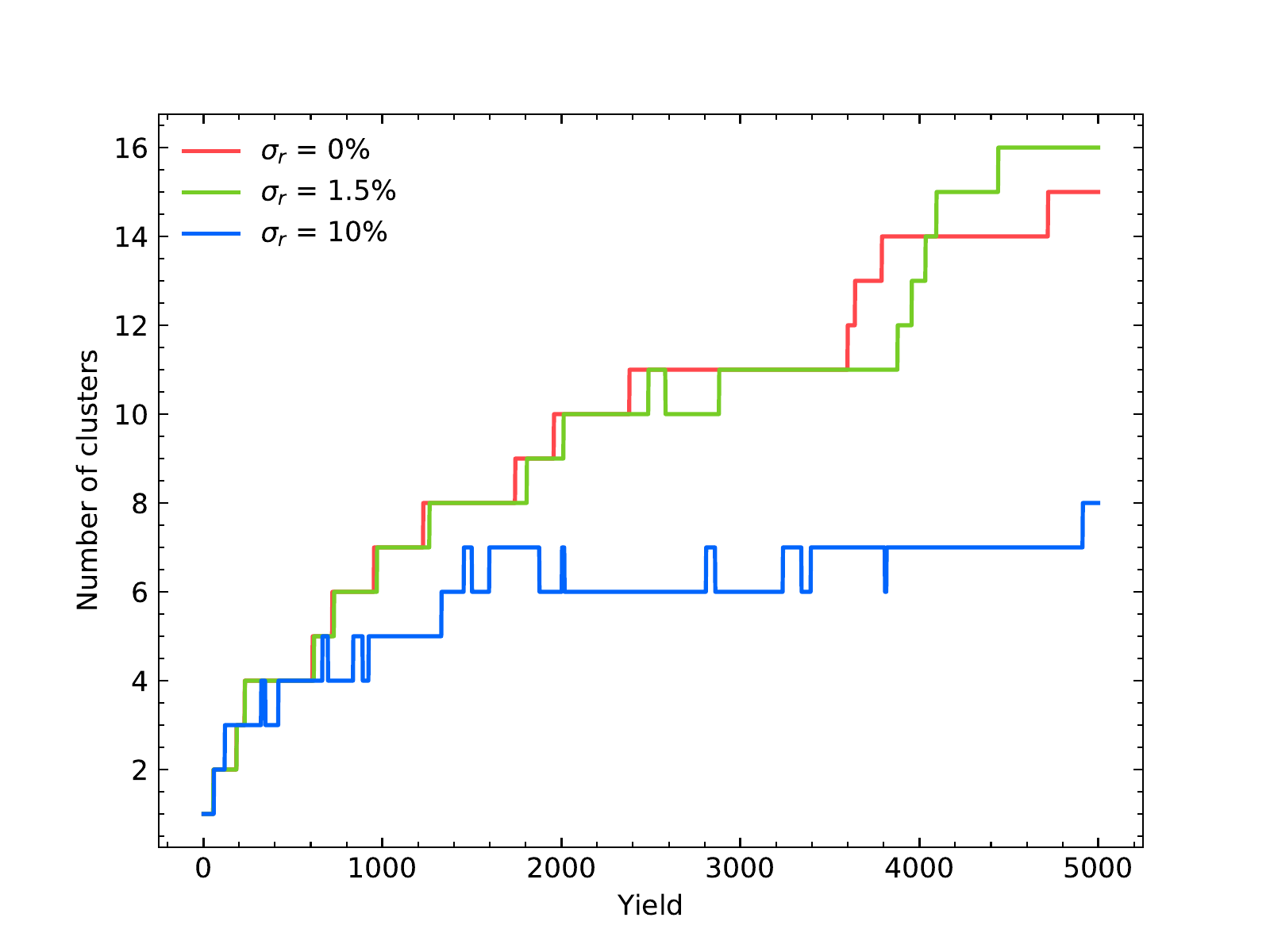}
\caption{The number of clusters as a function of the signal yield for the observable $\mathrm d\text{BR}(B\to D^0\tau \nu)/\mathrm dq^2$ for three different relative systematic uncertainties $\sigma_r$. }
\label{fig:clust_yield}
\end{figure}

The $q^2$-distribution of $\text{BR}(B\to D^0 \tau\nu)$ has already been studied extensively in the literature.
For our purpose we consider this observable to study the influence of the systematic and statistical uncertainties on the resulting number of clusters.
This is relevant for future experiments such as Belle II, reaching a higher luminosity than previously attained.
The $q^2$-distributions were computed using \texttt{flavio} \cite{Straub:2018kue}.
In \reffig{fig:clust_yield} we show the number of clusters as a function of the signal yield for various systematic uncertainties.
As expected, the number of resulting clusters is dominated by the signal yield for small yields, while systematic uncertainties become the limiting factor for larger yields.
\subsubsection{$E_\ell$-distribution of $\Gamma(\bar B\to D \tau^-(\to\ell\, \bar\nu_\ell\,\nu_\tau)\bar\nu_\tau)$}
Finally, we consider $E_\ell$, the energy of the light lepton $\ell$ from the 5-body decay $\bar B\to D \tau^-(\to\ell\, \bar\nu_\ell\,\nu_\tau)\bar\nu_\tau$.
To generate the kinematic distributions in terms of the chosen set of Wilson coefficients, we use the explicit expressions given in \cite{Alonso:2016gym} as further outlined in App.~\ref{app:Diffdist}.

The study of this observable is motivated by the BaBar analysis in \cite{Lees:2012xj}, where the experimental value of $R(D)$ was extracted under the assumption of a two-Higgs-doublet model.
Signal and background yields were extracted with a fit to the two-dimensional $m_{\text{miss}}^2, |\pell|$ distribution.
Here, $m_{\text{miss}}$ and $\pell$ denote the mass of the undetected neutrinos and the three-momentum of the light lepton in the rest frame of the $B$ meson.
The shape of these distributions for signal and background were taken from MC simulations.
Such MC simulations are usually assumed to be SM like.
In \cite{Lees:2012xj} however, the (SM) MC simulations were reweighted assuming non-zero values for the parameter $p\equiv \tan{\beta}/m_{H^\pm}^2$ of the two-Higgs-doublet model of type II\footnote{$\tan{\beta}$ denotes the ratio of the vacuum expectation values of the two Higgs bosons and $m_{H^\pm}$ is the mass of the charged Higgs boson.}.
The (model-dependent) experimental values of $R(D)$ were then extracted for 20 different values of $p$.
The resulting distribution of $R(D)$ measurements shows a sharp transition at $p= 0.35\,\mathrm{GeV}^{-1}$, but is otherwise relatively independent of $p$.

This result motivates to perform a clustering analysis to investigate the model dependency of the input kinematics.

\begin{figure}[tbp]
\centering
{
\begin{minipage}{12cm}
\centering
{\includegraphics[clip, trim=4.5cm 1cm 21cm 2.0cm, height=1.6cm]{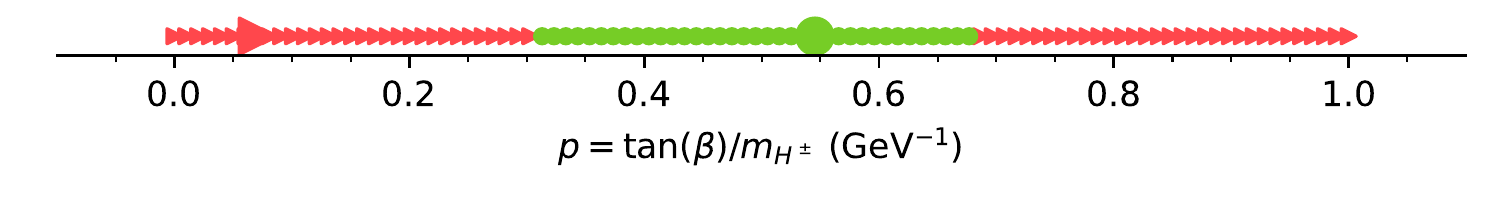}}
{\includegraphics[clip, trim=4.5cm 1cm 21cm 2.0cm, height=1.6cm]{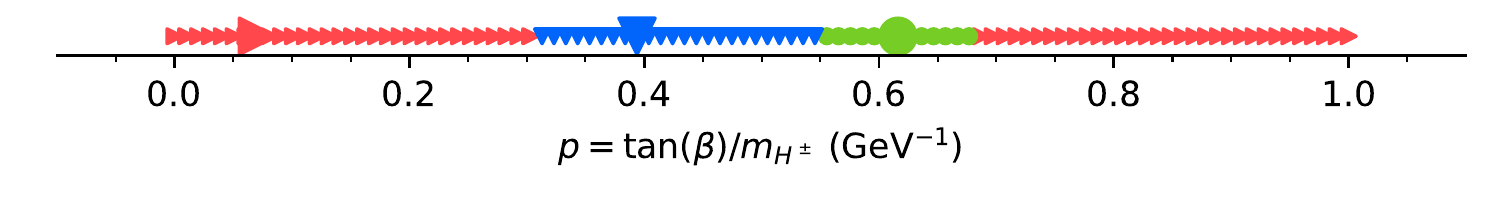}}
{\includegraphics[clip, trim=4.5cm 1cm 21cm 2.0cm, height=1.6cm]{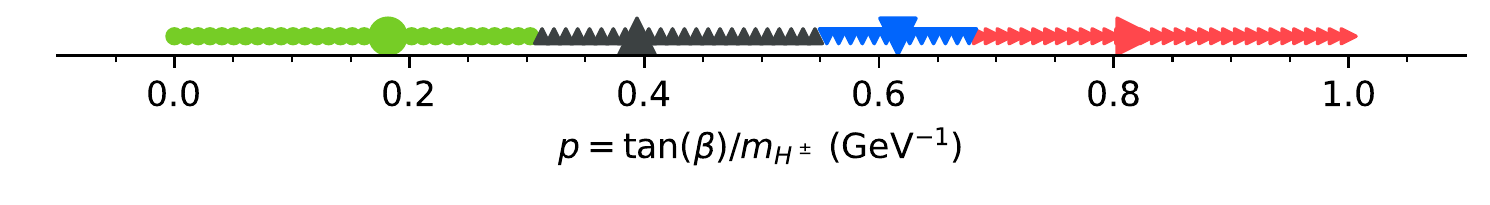}}
\end{minipage}}
\begin{minipage}{2cm}
\vspace{2cm}
\includegraphics[height=1.7cm]{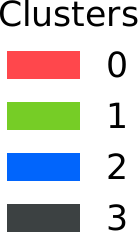}
\end{minipage}
\caption{Clustering of the one-dimensional parameter space $p$ of the observable $\mathrm d\Gamma(\bar B\to D \tau^-(\to\ell\, \bar\nu_\ell\,\nu_\tau)\bar\nu_\tau)/\mathrm d(E_\ell)$, with $E_\ell$ denoting the lepton energy.
The NP parameter $p$ stems from a two-Higgs-doublet model.
The clustering is performed assuming signal yields of 1000, 1800 and 2000 events as well as a relative uncertainty of 1\% and leads to a total of two, three and four clusters respectively.}
\label{fig:El_tanbeta_err}
\end{figure}

In this paper, we consider the $E_\ell$-distribution, which can be taken as an approximation of $|\pell|$ as considered in \cite{Lees:2012xj}.
The results of our clustering analysis with respect to the parameter $p$ are shown in \reffig{fig:El_tanbeta_err}.
The one-dimensional parameter space is clustered three times, assuming signal yields of 1000, 1800 and 2000 events as well as a relative systematic uncertainty of 1\%.
The first sub-figure shows the two resulting clusters for a yield of 1000 events, which coincide with the findings of \cite{Lees:2012xj}, where two different values of $R(D)$ are obtained.
However, in \cite{Lees:2012xj} the first value for $R(D)$ is obtained for $p\leq0.3$ and the second one for $p\geq 0.45$, whereas the first sub-figure suggests to have the same $R(D)$ for $0.3\leq p\leq 0.7$ and another value for the rest of the parameter space.

\begin{figure}[ht]
\begin{center}
\includegraphics[height=6.8cm, clip, trim=0cm 0cm 1.2cm 1cm]{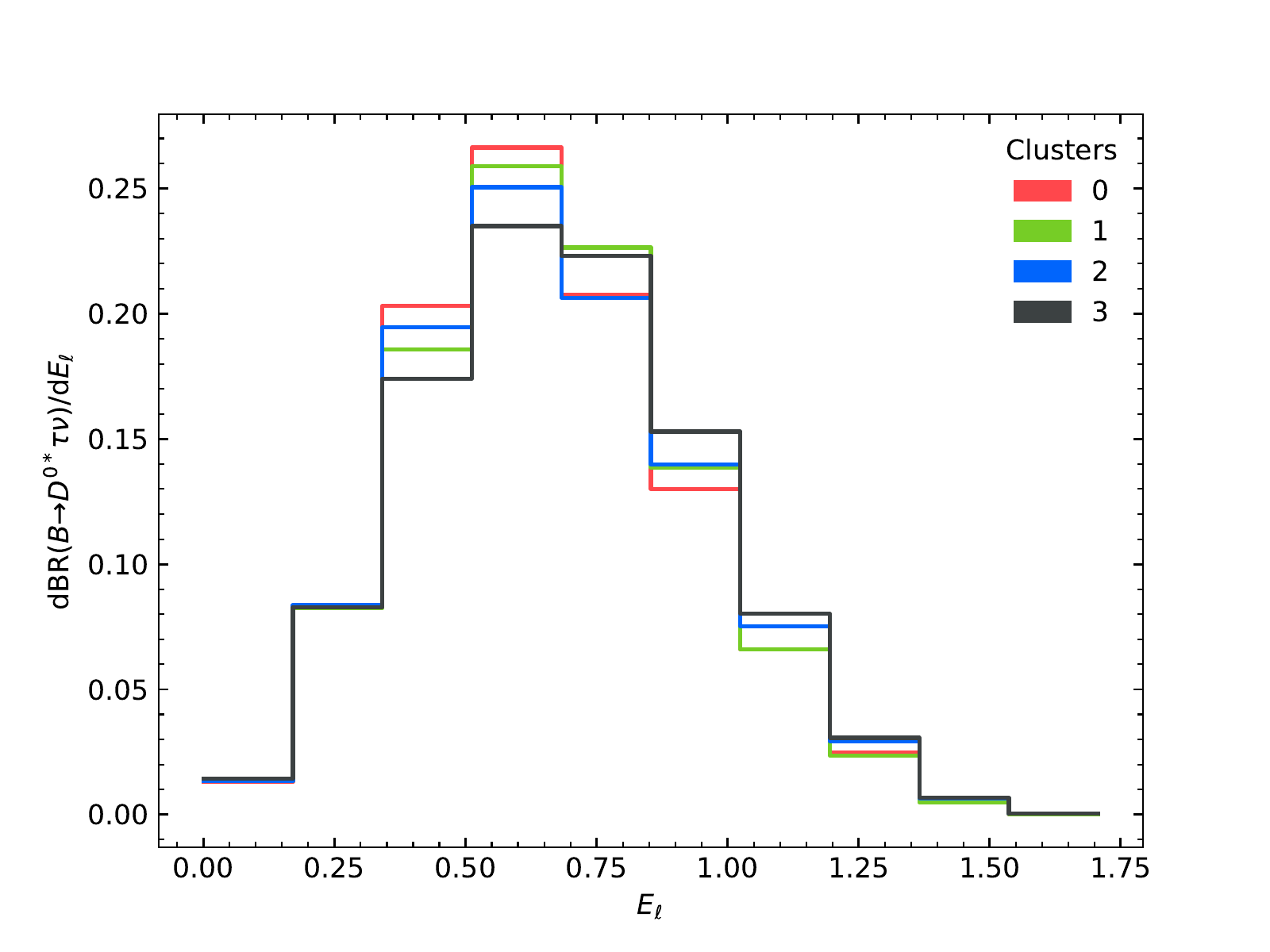}
\includegraphics[height=6.8cm, clip, trim=2.7cm 0cm 2cm 0cm]{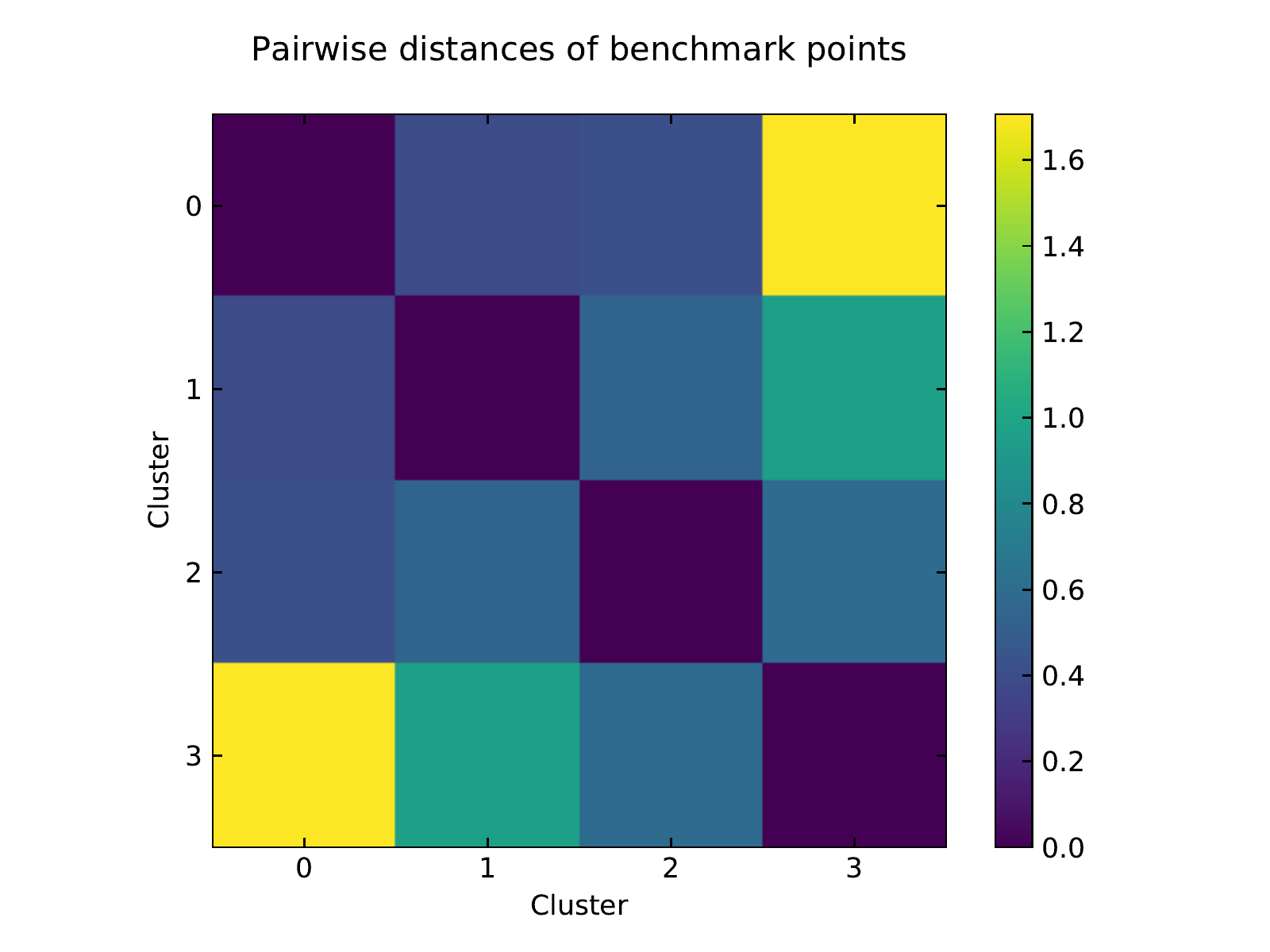}
\caption{Benchmark $E_\ell$-distributions of the decay width $\Gamma(\bar B\to D \tau^-(\to\ell\, \bar\nu_\ell\,\nu_\tau)\bar\nu_\tau)$ for a yield of 2000 events and a relative uncertainty of 1\%. The matrix plot shows the pairwise distances between the kinematic distributions.}
\label{fig:El_tanbeta_dist}
\end{center}
\end{figure}

Increasing the yield (and thereby reducing the uncertainties on the distributions) results in more clusters in the middle region (see \reffig{fig:El_tanbeta_err}), again indicating that the shape of the kinematic distribution significantly changes between $0.3 < p < 0.7$.
However, as can be seen from \reffig{fig:El_tanbeta_dist}, the kinematics for low and high $p$ are still very similar, incompatible with the result of \cite{Lees:2012xj}.

However, since \cite{Lees:2012xj} used the $m_{\text{miss}}^2$ distribution together with the $|\pell|$ distribution, it is not too surprising to arrive at a rather different result, as the shape of the distributions in $m_{\text{miss}}^2$ could behave very differently than $|\pell|$. Applying clustering techniques to the 2D $m_{\text{miss}}^2, |\pell|$ distribution will allow for a more thorough comparison and is left for future work.
\section{Conclusions}\label{sec:conclusions}
In this article we discussed cluster analyses of kinematic distributions.
These analyses divide the parameter space governing the distribution into subsets (\emph{clusters}), in which all points correspond to similar kinematic distributions.
Each cluster is then reduced to its most representative point (\emph{benchmark point}, \emph{BP}).
Analyses relying on these kinematic distributions can then be carried out for the BPs only, rather than using the full parameter space.
This can drastically reduce the required computing power and makes it easier to present numerical results as well as visualisations.

The results of the cluster analyses depend on the sampling of the parameter space, the clustering algorithm and the metric measuring differences between kinematic distributions.

This paper introduced the Python package \ck{} which implements the above steps and allows to perform clustering analyses without technical overhead.
While it particularly aims to make clustering techniques more accessible for the High Energy Physics community, the software can also be applied to more general problems outside of particle physics.
\ck{} is available at \url{github.com/clusterking} together with usage examples and technical documentation.

We used the \ck{} package to study several kinematic distributions of the decays $\bar B\to D^{0(*)}\tau^-\bar\nu_\tau$.
The $\theta_\tau$ and $\theta_V$ distribution of $\bar B\to D^{0*}\tau^-\bar\nu_\tau$ were studied, showing the clustered parameter space, the BPs as well as the corresponding distributions.
A strong dependence of the $\theta_\tau$-distribution on the tensor Wilson coefficient $C_T$ has been shown, which agrees with previous findings in the literature \cite{Jung:2018lfu}.

The influence of statistical and systematic uncertainties on the clustering result is shown on the example of the $q^2$-distribution of $\bar B\to D^{0}\tau^-\bar\nu_\tau$.

Finally we analysed the $E_\ell$-distribution of the 5-body decay $\bar B\to D \tau^-(\to\ell\, \bar\nu_\ell\,\nu_\tau)\bar\nu_\tau$.
The shape of this variable is an important input for some experimental measurements of $R(D)$.
The resulting model dependency that was observed in \cite{Lees:2012xj} on the example of type II two-Higgs-doublet models should also be seen from clustering the input kinematics.
While not entirely consistent with the results of \cite{Lees:2012xj}, our simplified approach correctly hints at the significant change of $R(D)$ at $\tan\beta/m^2_{H^\pm}\approx 0.3\,\mathrm{GeV}^{-1}$.
However, a full analysis including also the shape of the $m_{\text{miss}}^2$ distributions remains to be done in order to fully compare the results with those of \cite{Lees:2012xj}.
\section*{Acknowledgements}
We thank Alejandro Celis and David Straub for useful discussions.
The work of J.\,A., T.\,K. and K.\,L. is supported by the DFG clusters of excellence
\enquote{Origin and Structure of the Universe} and \enquote{ORIGINS: From the Origin of the Universe to the First Building Blocks of Life}.
\newpage
\appendix
\section{ClusterKinG example}\label{app:ck}
The following code example shows a full \ck{} workflow: Kinematic distributions for $\mathrm d\mathrm{BR} (B^+\rightarrow D^0\tau^-\bar\nu_\tau)/\mathrm d q^2$ are generated for a selection of sample points in Wilson space.
Uncertainties are added to the data and the distributions are clustered. Finally BPs are selected and several plots are generated.
Similar (and more systematic) examples can be found in the \texttt{examples} folder of the \ck{} repository.

\begin{lstlisting}[language=iPython]
import flavio
import numpy as np
import clusterking as ck
from clusterking.maths.metric import chi2_metric


# Define kinematic function using the flavio package
def dBrdq2(w, q):
    return flavio.np_prediction("dBR/dq2(B+->Dtaunu)", w, q)


# Set up and configure Scanner
s = ck.scan.WilsonScanner(scale=5, eft="WET", basis="flavio")
# Set kinematic function
s.set_dfunction(
    dBrdq2,
    binning=np.linspace(3.2, 11.6, 10),
    normalize=True
)
# Set sampling points in Wilson space
s.set_spoints_equidist({
    "CVL_bctaunutau": (-0.5, 0.5, 10),
    "CSL_bctaunutau": (-0.5, 0.5, 10),
    "CT_bctaunutau": (-0.1, 0.1, 10)
})

# Run scanner and add errors
d = ck.DataWithErrors()    # Create data object to write results to
r = s.run(d)               # Run scanner
r.write()                  # Write results back to data object
d.add_err_poisson(1000)    # statistical uncertainties
d.add_rel_err_uncorr(0.1)  # 0.1% relative system uncertainties, uncorrelated

# Clustering
c = ck.cluster.HierarchyCluster()  # Initialize worker class
c.set_metric(chi2_metric)
c.set_max_d(1)   # "Cut off" value for hierarchy
r = c.run(d)     # Run clustering on d
r.write()        # Write results back to data object

# Benchmarking
b = ck.Benchmark()  # Initialize worker class
b.set_metric(chi2_metric)
r = b.run(d)        # Run benchmarking
r.write()           # Write results back to data object

# Optional: Save data (kinematic distributions, clusters, BPs, ...)
d.write("btaunu_q2.sql")

# Find closest benchmark point to new parameter point
# Similar function for spoints: find_closest_spoints
d.find_closest_bpoints(
	{
		"CVL_bctaunutau": 0.1,
		"CSL_bctaunutau": 0.1,
		"CT_bctaunutau": 0.1,
	},
	n=1
)

# Generate plots similar to figures 1, 2, 3, 4, 7 respectively
d.plot_clusters_scatter(["CVL_bctaunutau", "CSL_bctaunutau"])
d.plot_dist_box()
d.plot_clusters_scatter(["CT_bctaunutau", "CVL_bctaunutau", "CSL_bctaunutau"])
d.plot_dist()
d.plot_bpoint_distance_matrix()
\end{lstlisting}
%
\section{Differential decay rates}\label{app:Diffdist}
The differential decay rates for the decays $\bar B\to D^{0(*)}\tau^-\bar\nu_\tau$ are given by:
\begin{align}
\frac{\mathrm d^2\Gamma(B\to D^{0(*)}\tau\nu)}{\mathrm dq^2\, \mathrm d(\cos{\theta_\ell})} &= \frac{3}{8}(I_1^c+2I_1^s)+\cos(\theta_\ell)\frac{3}{8}(I_6^c+2I_6^s)+\cos(2\theta_\ell)\frac{3}{8}(I_2^c+2I_2^s) \,,\\
  \frac{\mathrm d^2\Gamma(B\to D^{0*}\tau\nu)}{\mathrm dq^2\,\mathrm d(\cos{\theta_V})} &= -\cos^2(\theta_V)\frac{3}{8}(-3I_1^c+I_2^c)-\sin^2(\theta_V)\frac{3}{8}(-3I_1^s+I_2^s) \,, \\
  \frac{\mathrm d\Gamma(B\to D^{0*}\tau\nu)}{\mathrm dq^2} &= \frac{3}{4}(2I_1^s+I_1^c)-\frac{1}{4}(2I_2^s+I_2^c)\,,
\end{align}
where $q^2$ is the invariant mass of the lepton pair and $I_i^j$ denote the angular coefficients, which depend on $q^2$ and the Wilson coefficients.
The angle $\theta_\ell$ is defined as the angle between the direction of the tau in the dilepton rest frame and the direction of the dilepton in the $B$ rest frame, whereas $\theta_V$ is defined as the angle between the direction of the $D^{0*}$ meson in the dilepton rest frame and the direction of the $D^{0*}$ in the $B$ rest frame.
A general discussion of the kinematics of semileptonic meson decays (the context of $B\to K^*\bar\ell\ell$) together with the angular coefficients can be found in \cite{Gratrex:2015hna}.
For $\bar B\to D^{0(*)}\tau^-\bar\nu_\tau$ decays the corresponding distributions are discussed for example in \cite{Becirevic:2016hea,Alonso:2016gym}.
Our implementation of the above decay rates can be found at \href{https://github.com/clusterking/clusterking_physics/}{\texttt{github.com/clusterking/clusterking\_physics}}.
\section{Validation of statistical treatment}
\label{appendix:validation}
The statistical treatment of the examples shown in \refsec{sec:numerics} is validated with toy experiments: For each point in parameter space we consider the corresponding histogram and its covariance matrix.
Toy histograms are generated by drawing random values from the multivariate normal distribution with matching means and covariance matrix.
We then calculate the test statistic $\chi^2$ from \eqref{chi2_new} between each toy histogram and the original histogram.
The distribution of all $\chi^2/(N-1)$ values is binned and compared to the calculated expected distribution $\chi^2_{N-1}/(N-1)$ using the the Jenson-Shannon Divergence (JSD).

\begin{figure}[tb]
	\centering
	\includegraphics[scale=0.65]{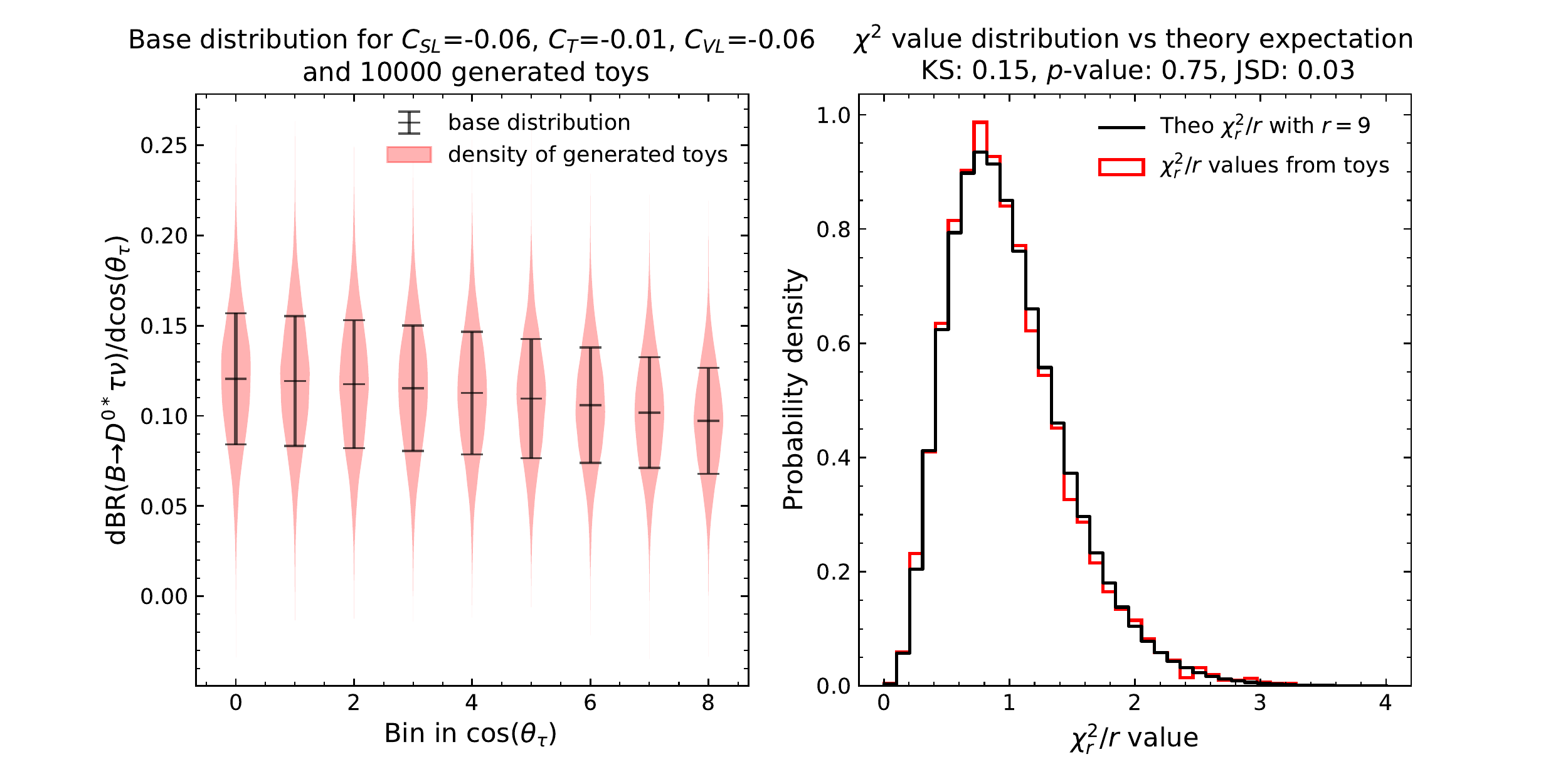}
	\caption{Toy experiments to validate the implementation of the $\chi^2$ metric. The right sided figure also reports several values quantifying the similarity of the toy distribution to the theoretical expectation: the Kolmogorov-Smirnov test statistic (KS), its corresponding $p$ value and the Jensen-Shannon Divergence (JSD).}
	\label{toy_experiment_demo}
\end{figure}

\begin{figure}[tb]
	\centering
	\includegraphics[scale=0.66]{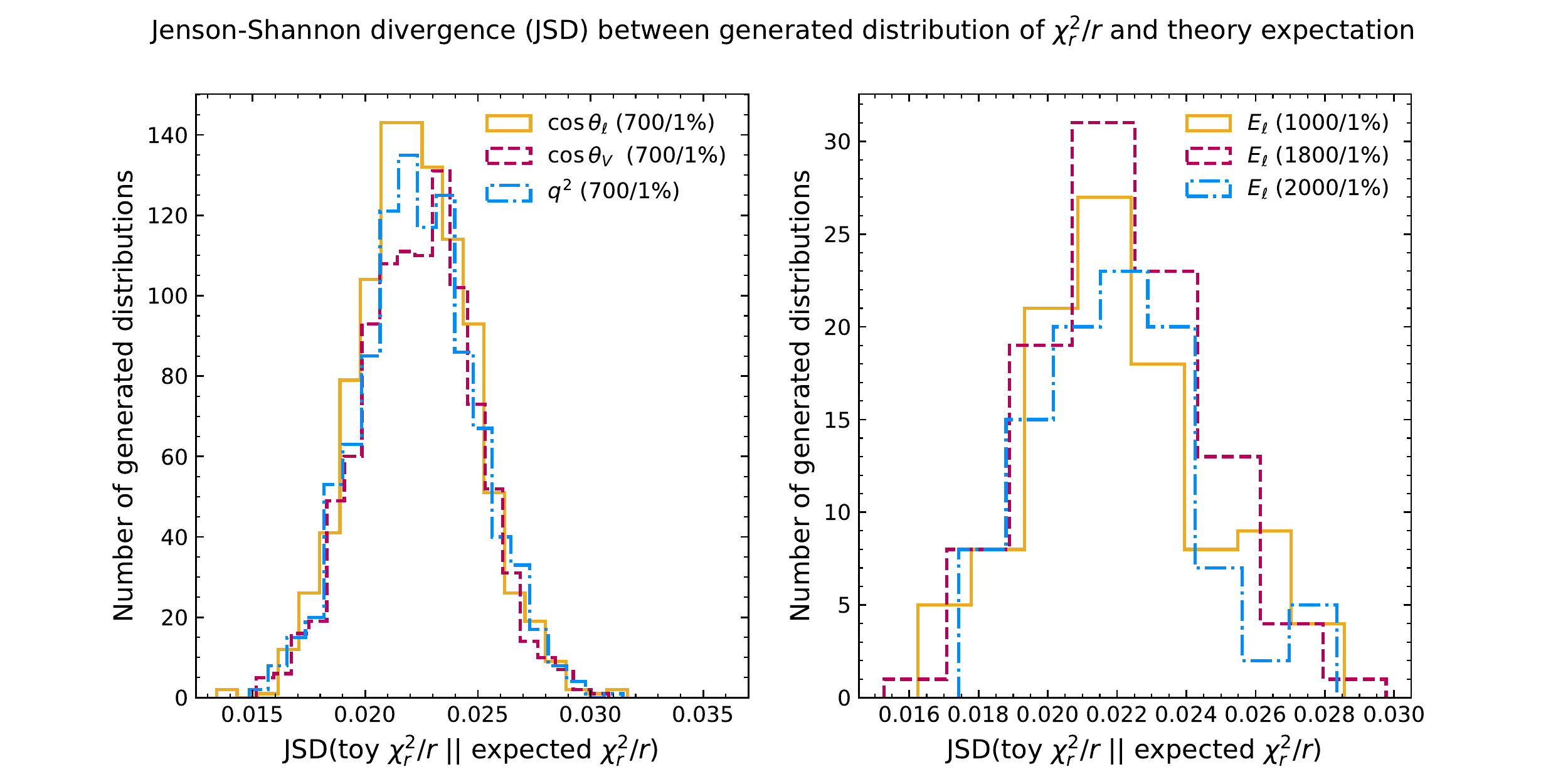}
	\caption{Validating the shape of the $\chi^2$ distribution for all points in the parameter space. The numbers in parentheses denote the assumed total yield corresponding to the Poisson uncertainties and the uncorrelated systematic uncertainty.}
	\label{toy_experiment_demo_all_points}
\end{figure}

An example for one particular point in parameter space is shown in \reffig{toy_experiment_demo}.
Both histograms agree nicely, resulting in a low JSD value.
The result of repeating the same procedure across all points is shown in \reffig{toy_experiment_demo_all_points}, showing satisfactorily low divergence values.

Additional code to reproduce the figures shown here and to validate the statistical treatment has been added to the \texttt{ClusterKinG} repository.

\bibliographystyle{JHEP}
\bibliography{bibliography}

\providecommand{\href}[2]{#2}\begingroup\raggedright\begin{thebibliography}{10}

\bibitem{Lees:2012xj}
\textbf{BaBar} Collaboration, J.~P. Lees et~al., \textit{{Evidence for an
  excess of $\bar{B} \to D^{(*)} \tau^-\bar{\nu}_\tau$ decays}},  {\em Phys.
  Rev. Lett.} \textbf{109} (2012) 101802,
  [\href{https://arxiv.org/abs/1205.5442}{\texttt{arXiv:1205.5442}}].

\bibitem{Carvalho:2015ttv}
A.~Carvalho, M.~Dall'Osso, T.~Dorigo, F.~Goertz, C.~A. Gottardo, and M.~Tosi,
  \textit{{Higgs Pair Production: Choosing Benchmarks With Cluster Analysis}},
  {\em JHEP} \textbf{04} (2016) 126,
  [\href{https://arxiv.org/abs/1507.02245}{\texttt{arXiv:1507.02245}}].

\bibitem{Carvalho:2016rys}
A.~Carvalho, M.~Dall'Osso, P.~De~Castro~Manzano, T.~Dorigo, F.~Goertz,
  M.~Gouzevich, and M.~Tosi, \textit{{Analytical parametrization and shape
  classification of anomalous HH production in the EFT approach}},
  \href{https://arxiv.org/abs/1608.06578}{\texttt{arXiv:1608.06578}}.

\bibitem{Carvalho:2017vnu}
A.~Carvalho, F.~Goertz, K.~Mimasu, M.~Gouzevitch, and A.~Aggarwal, \textit{{On
  the reinterpretation of non-resonant searches for Higgs boson pairs}},
  \href{https://arxiv.org/abs/1710.08261}{\texttt{arXiv:1710.08261}}.

\bibitem{Khachatryan:2016sey}
\textbf{CMS} Collaboration, V.~Khachatryan et~al., \textit{{Search for two
  Higgs bosons in final states containing two photons and two bottom quarks in
  proton-proton collisions at 8 TeV}},  {\em Phys. Rev.} \textbf{D94} (2016),
  no.~5 052012,
  [\href{https://arxiv.org/abs/1603.06896}{\texttt{arXiv:1603.06896}}].

\bibitem{Grzadkowski:2010es}
B.~Grzadkowski, M.~Iskrzynski, M.~Misiak, and J.~Rosiek, \textit{{Dimension-Six
  Terms in the Standard Model Lagrangian}},  {\em JHEP} \textbf{10} (2010) 085,
  [\href{https://arxiv.org/abs/1008.4884}{\texttt{arXiv:1008.4884}}].

\bibitem{Jenkins:2013zja}
E.~E. Jenkins, A.~V. Manohar, and M.~Trott, \textit{{Renormalization Group
  Evolution of the Standard Model Dimension Six Operators I: Formalism and
  lambda Dependence}},  {\em JHEP} \textbf{10} (2013) 087,
  [\href{https://arxiv.org/abs/1308.2627}{\texttt{arXiv:1308.2627}}].

\bibitem{Jenkins:2013wua}
E.~E. Jenkins, A.~V. Manohar, and M.~Trott, \textit{{Renormalization Group
  Evolution of the Standard Model Dimension Six Operators II: Yukawa
  Dependence}},  {\em JHEP} \textbf{01} (2014) 035,
  [\href{https://arxiv.org/abs/1310.4838}{\texttt{arXiv:1310.4838}}].

\bibitem{Alonso:2013hga}
R.~Alonso, E.~E. Jenkins, A.~V. Manohar, and M.~Trott, \textit{{Renormalization
  Group Evolution of the Standard Model Dimension Six Operators III: Gauge
  Coupling Dependence and Phenomenology}},  {\em JHEP} \textbf{04} (2014) 159,
  [\href{https://arxiv.org/abs/1312.2014}{\texttt{arXiv:1312.2014}}].

\bibitem{Celis:2017hod}
A.~Celis, J.~Fuentes-Martin, A.~Vicente, and J.~Virto, \textit{{DsixTools: The
  Standard Model Effective Field Theory Toolkit}},  {\em Eur. Phys. J.}
  \textbf{C77} (2017), no.~6 405,
  [\href{https://arxiv.org/abs/1704.04504}{\texttt{arXiv:1704.04504}}].

\bibitem{Aebischer:2015fzz}
J.~Aebischer, A.~Crivellin, M.~Fael, and C.~Greub, \textit{{Matching of gauge
  invariant dimension-six operators for $b\to s$ and $b\to c$ transitions}},
  {\em JHEP} \textbf{05} (2016) 037,
  [\href{https://arxiv.org/abs/1512.02830}{\texttt{arXiv:1512.02830}}].

\bibitem{Jenkins:2017jig}
E.~E. Jenkins, A.~V. Manohar, and P.~Stoffer, \textit{{Low-Energy Effective
  Field Theory below the Electroweak Scale: Operators and Matching}},  {\em
  JHEP} \textbf{03} (2018) 016,
  [\href{https://arxiv.org/abs/1709.04486}{\texttt{arXiv:1709.04486}}].

\bibitem{Dekens:2019ept}
W.~Dekens and P.~Stoffer, \textit{{Low-energy effective field theory below the
  electroweak scale: matching at one loop}},  {\em JHEP} \textbf{10} (2019)
  197, [\href{https://arxiv.org/abs/1908.05295}{\texttt{arXiv:1908.05295}}].

\bibitem{Aebischer:2017gaw}
J.~Aebischer, M.~Fael, C.~Greub, and J.~Virto, \textit{{B physics Beyond the
  Standard Model at One Loop: Complete Renormalization Group Evolution below
  the Electroweak Scale}},  {\em JHEP} \textbf{09} (2017) 158,
  [\href{https://arxiv.org/abs/1704.06639}{\texttt{arXiv:1704.06639}}].

\bibitem{Jenkins:2017dyc}
E.~E. Jenkins, A.~V. Manohar, and P.~Stoffer, \textit{{Low-Energy Effective
  Field Theory below the Electroweak Scale: Anomalous Dimensions}},  {\em JHEP}
  \textbf{01} (2018) 084,
  [\href{https://arxiv.org/abs/1711.05270}{\texttt{arXiv:1711.05270}}].

\bibitem{Criado:2017khh}
J.~C. Criado, \textit{{MatchingTools: a Python library for symbolic effective
  field theory calculations}},  {\em Comput. Phys. Commun.} \textbf{227} (2018)
  42--50, [\href{https://arxiv.org/abs/1710.06445}{\texttt{arXiv:1710.06445}}].

\bibitem{Aebischer:2018bkb}
J.~Aebischer, J.~Kumar, and D.~M. Straub, \textit{{Wilson: a Python package for
  the running and matching of Wilson coefficients above and below the
  electroweak scale}},  {\em Eur. Phys. J.} \textbf{C78} (2018), no.~12 1026,
  [\href{https://arxiv.org/abs/1804.05033}{\texttt{arXiv:1804.05033}}].

\bibitem{Aebischer:2017ugx}
J.~Aebischer et~al., \textit{{WCxf: an exchange format for Wilson coefficients
  beyond the Standard Model}},  {\em Comput. Phys. Commun.} \textbf{232} (2018)
  71--83, [\href{https://arxiv.org/abs/1712.05298}{\texttt{arXiv:1712.05298}}].

\bibitem{Aebischer:2018iyb}
J.~Aebischer, J.~Kumar, P.~Stangl, and D.~M. Straub, \textit{{A Global
  Likelihood for Precision Constraints and Flavour Anomalies}},  {\em Eur.
  Phys. J.} \textbf{C79} (2019), no.~6 509,
  [\href{https://arxiv.org/abs/1810.07698}{\texttt{arXiv:1810.07698}}].

\bibitem{Criado:2019ugp}
J.~C. Criado, \textit{{BasisGen: automatic generation of operator bases}},
  {\em Eur. Phys. J.} \textbf{C79} (2019), no.~3 256,
  [\href{https://arxiv.org/abs/1901.03501}{\texttt{arXiv:1901.03501}}].

\bibitem{Gripaios:2018zrz}
B.~Gripaios and D.~Sutherland, \textit{{DEFT: A program for operators in EFT}},
   {\em JHEP} \textbf{01} (2019) 128,
  [\href{https://arxiv.org/abs/1807.07546}{\texttt{arXiv:1807.07546}}].

\bibitem{Brivio:2019irc}
I.~Brivio et~al., \textit{{Computing Tools for the SMEFT}},  in {\em {Computing
  Tools for the SMEFT}} (J.~Aebischer, M.~Fael, A.~Lenz, M.~Spannowsky, and
  J.~Virto, eds.), 2019.
\newblock \href{https://arxiv.org/abs/1910.11003}{\texttt{arXiv:1910.11003}}.

\bibitem{Straub:2018kue}
D.~M. Straub, \textit{{flavio: a Python package for flavour and precision
  phenomenology in the Standard Model and beyond}},
  \href{https://arxiv.org/abs/1810.08132}{\texttt{arXiv:1810.08132}}.

\bibitem{Porod:2014xia}
W.~Porod, F.~Staub, and A.~Vicente, \textit{{A Flavor Kit for BSM models}},
  {\em Eur. Phys. J.} \textbf{C74} (2014), no.~8 2992,
  [\href{https://arxiv.org/abs/1405.1434}{\texttt{arXiv:1405.1434}}].

\bibitem{Mahmoudi:2007vz}
F.~Mahmoudi, \textit{{SuperIso: A Program for calculating the isospin asymmetry
  of $B \rightarrow K^* \gamma$ in the MSSM}},  {\em Comput. Phys. Commun.}
  \textbf{178} (2008) 745--754,
  [\href{https://arxiv.org/abs/0710.2067}{\texttt{arXiv:0710.2067}}].

\bibitem{Porod:2003um}
W.~Porod, \textit{{SPheno, a program for calculating supersymmetric spectra,
  SUSY particle decays and SUSY particle production at e+ e- colliders}},  {\em
  Comput. Phys. Commun.} \textbf{153} (2003) 275--315,
  [\href{https://arxiv.org/abs/hep-ph/0301101}{\texttt{hep-ph/0301101}}].

\bibitem{Porod:2011nf}
W.~Porod and F.~Staub, \textit{{SPheno 3.1: Extensions including flavour,
  CP-phases and models beyond the MSSM}},  {\em Comput. Phys. Commun.}
  \textbf{183} (2012) 2458--2469,
  [\href{https://arxiv.org/abs/1104.1573}{\texttt{arXiv:1104.1573}}].

\bibitem{Evans:2016lzo}
J.~A. Evans and D.~Shih, \textit{{FormFlavor Manual}},
  \href{https://arxiv.org/abs/1606.00003}{\texttt{arXiv:1606.00003}}.

\bibitem{Dedes:2019uzs}
A.~Dedes, M.~Paraskevas, J.~Rosiek, K.~Suxho, and L.~Trifyllis,
  \textit{{SmeftFR - Feynman rules generator for the Standard Model Effective
  Field Theory}},
  \href{https://arxiv.org/abs/1904.03204}{\texttt{arXiv:1904.03204}}.

\bibitem{Brivio:2017btx}
I.~Brivio, Y.~Jiang, and M.~Trott, \textit{{The SMEFTsim package, theory and
  tools}},  {\em JHEP} \textbf{12} (2017) 070,
  [\href{https://arxiv.org/abs/1709.06492}{\texttt{arXiv:1709.06492}}].

\bibitem{Altmannshofer:2017poe}
W.~Altmannshofer, P.~S. Bhupal~Dev, and A.~Soni, \textit{{$R_{D^{(*)}}$
  anomaly: A possible hint for natural supersymmetry with $R$-parity
  violation}},  {\em Phys. Rev.} \textbf{D96} (2017), no.~9 095010,
  [\href{https://arxiv.org/abs/1704.06659}{\texttt{arXiv:1704.06659}}].

\bibitem{Murgui:2019czp}
C.~Murgui, A.~Peñuelas, M.~Jung, and A.~Pich, \textit{{Global fit to $b \to c
  \tau \nu$ transitions}},
  \href{https://arxiv.org/abs/1904.09311}{\texttt{arXiv:1904.09311}}.

\bibitem{Shi:2019gxi}
R.-X. Shi, L.-S. Geng, B.~Grinstein, S.~Jäger, and J.~Martin~Camalich,
  \textit{{Revisiting the new-physics interpretation of the $b\to c\tau\nu$
  data}},  \href{https://arxiv.org/abs/1905.08498}{\texttt{arXiv:1905.08498}}.

\bibitem{Becirevic:2019tpx}
D.~Bečirević, M.~Fedele, I.~Nišandžić, and A.~Tayduganov, \textit{{Lepton
  Flavor Universality tests through angular observables of $\overline{B}\to
  D^{(\ast)}\ell\overline{\nu}$ decay modes}},
  \href{https://arxiv.org/abs/1907.02257}{\texttt{arXiv:1907.02257}}.

\bibitem{Gomez:2019xfw}
J.~D. Gómez, N.~Quintero, and E.~Rojas, \textit{{Charged current $b \to c \tau
  \bar{\nu}_\tau$ anomalies in a general $W^\prime$ boson scenario}},  {\em
  Phys. Rev.} \textbf{D100} (2019), no.~9 093003,
  [\href{https://arxiv.org/abs/1907.08357}{\texttt{arXiv:1907.08357}}].

\bibitem{Alguero:2019ptt}
M.~Algueró, B.~Capdevila, A.~Crivellin, S.~Descotes-Genon, P.~Masjuan,
  J.~Matias, and J.~Virto, \textit{{Emerging patterns of New Physics with and
  without Lepton Flavour Universal contributions}},  {\em Eur. Phys. J.}
  \textbf{C79} (2019), no.~8 714,
  [\href{https://arxiv.org/abs/1903.09578}{\texttt{arXiv:1903.09578}}].

\bibitem{Ciuchini:2019usw}
M.~Ciuchini, A.~M. Coutinho, M.~Fedele, E.~Franco, A.~Paul, L.~Silvestrini, and
  M.~Valli, \textit{{New Physics in $b \to s \ell^+ \ell^-$ confronts new data
  on Lepton Universality}},  {\em Eur. Phys. J.} \textbf{C79} (2019), no.~8
  719, [\href{https://arxiv.org/abs/1903.09632}{\texttt{arXiv:1903.09632}}].

\bibitem{Datta:2019zca}
A.~Datta, J.~Kumar, and D.~London, \textit{{The $B$ anomalies and new physics
  in $b \to s e^+ e^-$}},  {\em Phys. Lett.} \textbf{B797} (2019) 134858,
  [\href{https://arxiv.org/abs/1903.10086}{\texttt{arXiv:1903.10086}}].

\bibitem{Aebischer:2019mlg}
J.~Aebischer, W.~Altmannshofer, D.~Guadagnoli, M.~Reboud, P.~Stangl, and D.~M.
  Straub, \textit{{$B$-decay discrepancies after Moriond 2019}},
  \href{https://arxiv.org/abs/1903.10434}{\texttt{arXiv:1903.10434}}.

\bibitem{Bernlochner:2020tfi}
F.~U. Bernlochner, S.~Duell, Z.~Ligeti, M.~Papucci, and D.~J. Robinson,
  \textit{{Das ist der HAMMER: Consistent new physics interpretations of
  semileptonic decays}},
  \href{https://arxiv.org/abs/2002.00020}{\texttt{arXiv:2002.00020}}.

\bibitem{Duell:2016maj}
S.~Duell, F.~Bernlochner, Z.~Ligeti, M.~Papucci, and D.~Robinson,
  \textit{{HAMMER: Reweighting tool for simulated data samples}},  {\em PoS}
  \textbf{ICHEP2016} (2017) 1074.

\bibitem{kolmogorov_1951}
F.~J. Massey, \textit{The {K}olmogorov-{S}mirnov test for goodness of fit},
  {\em Journal of the American Statistical Association} \textbf{46} (1951),
  no.~253 68--78.

\bibitem{10.1093/biomet/63.1.161}
A.~N. PETTITT, \textit{{A two-sample Anderson-Darling rank statistic}},  {\em
  Biometrika} \textbf{63} (04, 1976) 161--168,
  [\href{https://arxiv.org/abs/http://oup.prod.sis.lan/biomet/article-pdf/63/1/161/6689229/63-1-161.pdf}{\texttt{http://oup.prod.sis.lan/biomet/article-pdf/63/1/161/6689229/63-1-161.pdf}}].

\bibitem{wilks1938}
S.~S. Wilks, \textit{The large-sample distribution of the likelihood ratio for
  testing composite hypotheses},  {\em Ann. Math. Statist.} \textbf{9} (03,
  1938) 60--62.

\bibitem{Hartigan:105051}
J.~A. Hartigan, {\em {Clustering algorithms}}.
\newblock Wiley series in probability and mathematical statistics. Wiley, New
  York, NY, 1975.

\bibitem{macqueen1967}
J.~MacQueen, \textit{Some methods for classification and analysis of
  multivariate observations},  in {\em Proceedings of the Fifth Berkeley
  Symposium on Mathematical Statistics and Probability, Volume 1: Statistics},
  (Berkeley, Calif.), pp.~281--297, University of California Press, 1967.

\bibitem{Kaufman1990FindingGI}
L.~Kaufman and P.~Rousseeuw, {\em Finding Groups in Data: An Introduction to
  Cluster Analysis}.
\newblock John Wiley \& Sons, 1990.

\bibitem{mckinney-proc-scipy-2010}
W.~McKinney, \textit{Data structures for statistical computing in python},  in
  {\em Proceedings of the 9th Python in Science Conference} (S.~van~der Walt
  and J.~Millman, eds.), pp.~51 -- 56, 2010.

\bibitem{scipy}
E.~Jones, T.~Oliphant, P.~Peterson, et~al., \textit{{SciPy}: Open source
  scientific tools for {Python}},  2001.

\bibitem{Celis:2016azn}
A.~Celis, M.~Jung, X.-Q. Li, and A.~Pich, \textit{{Scalar contributions to
  $b\to c (u) \tau \nu$ transitions}},  {\em Phys. Lett.} \textbf{B771} (2017)
  168--179,
  [\href{https://arxiv.org/abs/1612.07757}{\texttt{arXiv:1612.07757}}].

\bibitem{Blanke:2018yud}
M.~Blanke, A.~Crivellin, S.~de~Boer, T.~Kitahara, M.~Moscati, U.~Nierste, and
  I.~Nišandžić, \textit{{Impact of polarization observables and $ B_c\to
  \tau \nu$ on new physics explanations of the $b\to c \tau \nu$ anomaly}},
  {\em Phys. Rev.} \textbf{D99} (2019), no.~7 075006,
  [\href{https://arxiv.org/abs/1811.09603}{\texttt{arXiv:1811.09603}}].

\bibitem{Bhattacharya:2018kig}
S.~Bhattacharya, S.~Nandi, and S.~Kumar~Patra, \textit{{$b \rightarrow c \tau
  \nu _{\tau }$ Decays: a catalogue to compare, constrain, and correlate new
  physics effects}},  {\em Eur. Phys. J.} \textbf{C79} (2019), no.~3 268,
  [\href{https://arxiv.org/abs/1805.08222}{\texttt{arXiv:1805.08222}}].

\bibitem{Becirevic:2018afm}
D.~Bečirević, I.~Doršner, S.~Fajfer, N.~Košnik, D.~A. Faroughy, and
  O.~Sumensari, \textit{{Scalar leptoquarks from grand unified theories to
  accommodate the $B$-physics anomalies}},  {\em Phys. Rev.} \textbf{D98}
  (2018), no.~5 055003,
  [\href{https://arxiv.org/abs/1806.05689}{\texttt{arXiv:1806.05689}}].

\bibitem{Bhattacharya:2019olg}
B.~Bhattacharya, A.~Datta, S.~Kamali, and D.~London, \textit{{CP Violation in
  ${\bar B}^0\to D^{*+}\mu^-{\bar\nu}_\mu$}},  {\em JHEP} \textbf{05} (2019)
  191, [\href{https://arxiv.org/abs/1903.02567}{\texttt{arXiv:1903.02567}}].

\bibitem{Adamczyk:2019wyt}
\textbf{Belle, Belle-II} Collaboration, K.~Adamczyk, \textit{{Semitauonic $B$
  decays at Belle/Belle II}},  in {\em {10th International Workshop on the CKM
  Unitarity Triangle (CKM 2018) Heidelberg, Germany, September 17-21, 2018}},
  2019.
\newblock \href{https://arxiv.org/abs/1901.06380}{\texttt{arXiv:1901.06380}}.

\bibitem{Jung:2018lfu}
M.~Jung and D.~M. Straub, \textit{{Constraining new physics in $b\to c\ell\nu$
  transitions}},  {\em JHEP} \textbf{01} (2019) 009,
  [\href{https://arxiv.org/abs/1801.01112}{\texttt{arXiv:1801.01112}}].

\bibitem{Alonso:2016gym}
R.~Alonso, A.~Kobach, and J.~Martin~Camalich, \textit{{New physics in the
  kinematic distributions of $\bar B\to
  D^{(*)}\tau^-(\to\ell^-\bar\nu_\ell\nu_\tau)\bar\nu_\tau$}},  {\em Phys.
  Rev.} \textbf{D94} (2016), no.~9 094021,
  [\href{https://arxiv.org/abs/1602.07671}{\texttt{arXiv:1602.07671}}].

\bibitem{Gratrex:2015hna}
J.~Gratrex, M.~Hopfer, and R.~Zwicky, \textit{{Generalised helicity formalism,
  higher moments and the $B \to K_{J_K}(\to K \pi) \bar{\ell}_1 \ell_2$ angular
  distributions}},  {\em Phys. Rev.} \textbf{D93} (2016), no.~5 054008,
  [\href{https://arxiv.org/abs/1506.03970}{\texttt{arXiv:1506.03970}}].

\bibitem{Becirevic:2016hea}
D.~Becirevic, S.~Fajfer, I.~Nisandzic, and A.~Tayduganov, \textit{{Angular
  distributions of $\bar B \to D^{(\ast)}\ell\bar \nu_\ell$ decays and search
  of New Physics}},
  \href{https://arxiv.org/abs/1602.03030}{\texttt{arXiv:1602.03030}}.

\end{thebibliography}\endgroup
\end{document}